\documentclass[pdflatex,sn-nature]{sn-jnl}

\usepackage[utf8]{inputenc}
\usepackage[T1]{fontenc}
\usepackage{lmodern}
\usepackage{microtype}
\usepackage[english]{babel}
\usepackage{bookmark}

\usepackage{manyfoot}
\usepackage{amsthm}
\usepackage{textcomp}

\usepackage{acro}
\DeclareAcronym{cct}{short = CCT, long = critical clearing time}
\DeclareAcronym{mae}{short = MAE, long = mean absolute error}
\DeclareAcronym{rmse}{short = RMSE, long = root-mean-square error}
\DeclareAcronym{rms}{short = RMS, long = root-mean-square}
\DeclareAcronym{naurc}{short = nAURC, long = normalized area under the risk-coverage curve}
\DeclareAcronym{hnsw}{short = HNSW, long = hierarchical navigable small world}

\usepackage{graphicx}
\graphicspath{{figures/}}
\usepackage{amsmath}
\usepackage{siunitx}
\usepackage{booktabs}
\usepackage{multirow}
\usepackage{csvsimple}
\usepackage{pgfplotstable}
\pgfplotsset{compat=1.18}
\usepgfplotslibrary{groupplots}
\usepackage{tikz}
\usetikzlibrary{shapes.geometric, positioning, arrows.meta}

\unnumbered

\begin{document}
\vbadness=10000

\title[Archive-constrained retrieval]{Archive coverage constrains abstaining retrieval for transient stability screening more than conditional accuracy}

\author*[1,2]{\fnm{Gregor} \sur{Cerar}}\email{gregor.cerar@ijs.si}
\author[1,2]{\fnm{Denis} \sur{Sodin}}
\author[3]{\fnm{Matjaž} \sur{Škrlec}}
\author*[3]{\fnm{Urban} \sur{Rudež}}\email{urban.rudez@fe.uni-lj.si}
\author[1]{\fnm{Mihael} \sur{Mohorčič}}
\author[1,2]{\fnm{Miha} \sur{Smolnikar}}

\affil*[1]{\orgname{Jožef Stefan Institute}, \orgaddress{\city{Ljubljana}, \country{Slovenia}}}
\affil[2]{\orgname{ComSensus d.o.o.}, \orgaddress{\city{Ljubljana}, \country{Slovenia}}}
\affil[3]{\orgname{University of Ljubljana, Faculty of Electrical Engineering}, \orgaddress{\city{Ljubljana}, \country{Slovenia}}}

\abstract{Power-grid transient stability is assessed by simulating credible faults, but repeating simulations as operating conditions change is costly, motivating faster critical clearing time estimators. Within the AI-ASSIST project, we developed a retrieval-based estimator around three adopted requirements: estimates retain their supporting states, retrieval abstains without compatible support, and support statistics are tested against observed error. Abstention turns estimator uncertainty into a demand on the archive. Exact switching-configuration equality is used. The dense synthetic BUS39 archive covers 99.9996\% of 1,044,797 records; the operational ELES archive covers 73.47\% of 322,823 records. At query time, 59.37\% of ELES states have an earlier same-topology candidate. Under causal availability, retrieval covers 56.34\% of records; on records covered under both protocols, mean absolute error increases by 10.0\%. No tested support diagnostic ranks first on both grids. Most tested supervised regressors have lower error than retrieval on BUS39, all higher on the matched retrieval-covered ELES population. On ELES, gradient boosting becomes more accurate than retrieval when temporally adjacent states are excluded, so the accuracy trade depends on the archive. Under the adopted compatibility key, archive availability constrains deployment more than conditional accuracy on ELES. Compatible coverage and chronological availability are auditable before an estimator is fitted.}


\maketitle

\section{Introduction}\label{sec:intro}

Transient stability describes whether a power grid's generators remain in synchronism after a large disturbance, such as a short circuit. It is assessed by simulating the grid's dynamic response to candidate faults and measuring the \acl{cct} (\acs{cct}), the longest fault duration for which the system still returns to synchronous operation once the fault is cleared~\cite{kundur2004definition}. Time-domain simulation is the reference method, but a full contingency screen may take longer than an operating state remains current~\cite{pavella2000transient,chiang2011direct}.

The shift from synchronous generation toward converter-interfaced wind and solar generation can reduce synchronous inertia where it displaces synchronous machines and introduce new stability and control challenges~\cite{evald2026gridforming}. Many data-driven methods use post-fault trajectories or phasor measurements~\cite{hu2019svm,fan2018knn,mazhari2021sbc,lv2023gedfscl}, as reviewed in~\cite{qinghua2023survey}. These methods incorporate information about the developing transient but cannot assess the pre-fault window. Other methods predict directly from steady-state pre-fault conditions~\cite{sulistiawati2016cct}, which is the setting considered here. Both groups commonly return one class label or scalar \acs{cct} for an assumed fault location, although supervised models can predict predefined sets of fault scenarios~\cite{chen2021preventive}. Such predictions do not identify the historical states underlying an estimate, and abstention has been studied separately from stability assessment~\cite{chow1970reject,elyaniv2010selective,geifman2017selective,guo2014probabilistic}.

A case-based alternative retrieves, for a new pre-fault state, historical or simulated states with similar conditions and aggregates their known outcomes~\cite{aamodt1994case}. Applied to transient stability, this follows earlier instance-based and clustering approaches~\cite{su2016clustering}, within a longer line of data-driven security assessment~\cite{wehenkel1989,morison2004security,xiao2020data} and recent work on making stability predictions interpretable~\cite{ren2022dl}. This literature does not establish how such a method should behave when its output informs operational decisions. Within the AI-ASSIST project, we developed a retrieval method to satisfy three requirements adopted for that setting. R1: an estimate must retain the recorded states that produced it and the rule that combined them, enabling independent audit. R2: the method must not extrapolate across switching configurations; where the archive contains no topology-compatible state, it must abstain rather than return a number. R3: statistics describing retrieval support must be evaluated against observed error rather than presented as measures of confidence.

R2 converts directly into a requirement on the archive: an estimate exists only where compatible states have already been simulated. It therefore raises the central question of whether an operational archive can supply that support, which we evaluate alongside R1 and R3. Sufficient coverage and acceptable referral load depend on the task and the cost of the simulation avoided; we therefore report the observed archive characteristics without defining a universal threshold. Computational performance is not evaluated.

These measurements differ in their determinants. Whether compatible states exist, and whether they existed when the query arrived, are archive properties under the adopted compatibility key. These properties can be evaluated before an estimator is fitted and bound any method that abstains outside topology-compatible support. The support retained after restricting retrieval to earlier states depends on both these properties and the eligibility rule, whereas the associated accuracy penalty is specific to this implementation.

The sections below evaluate the implementation and consequences of these requirements on a dense synthetic dataset generated on the public IEEE 39-bus network and on seven months of operational data from ELES, the Slovenian transmission system operator. R1 is enforced by construction: an estimate retains its supporting simulated cases, groups them by their recorded outcomes, and reports the aggregation rule. This representation also enables the critical-generator analysis below. R2 is evaluated through compatible coverage, chronological availability, support retained after causal reference selection, and referral workload. R3 is evaluated by testing whether diagnostics of retrieval support rank estimates according to observed error.

The principal empirical finding concerns R2: in the operational archive, compatible support becomes unavailable before conditional error increases substantially, whereas the synthetic archive provides nearly complete coverage.

The estimation task is conditional rather than predictive because the recorded pre-fault state does not determine where a fault will occur. Faults originate from weather, vegetation contact, lightning, and equipment failure, none of which are represented by the load-flow variables used here. A single retrieval estimates, for every location represented among its neighbors, whether the system would remain stable following a fault at that location, and abstains where no such location is represented. A supervised regressor must instead be provided with an assumed location and returns an estimate regardless of whether a comparable case exists. For the accuracy comparison, both methods receive the recorded location, thereby isolating the comparison from this difference.

\section{Results}\label{sec:results}

Four terms recur below and are specified in full in Methods. A state's switching configuration is the vector of line-status flags recorded with it; two states are compatible only when those vectors are equal, and nearly identical configurations are treated as incompatible. Exact equality is adopted to ensure that retrieved cases share the query's switching configuration, not because it improves predictive accuracy. Where no compatible state exists, the method abstains rather than returning a number; this is R2 and the basis of every coverage figure reported here. The critical generator is a simulation outcome rather than an input. \acs{cct} labels are resolved to \qty{0.01}{\second}, so no label separates clearing times closer than this resolution, and margins below one step should be interpreted accordingly.

BUS39 contains 1,044,797 state-contingency records from 21,783 states and ELES 322,823 records from 4,393 states. Both are evaluated leave-one-state-out, each query retrieving from every state but its own (Methods). A query returns the states sharing its switching configuration, grouped by the fault location and critical generator recorded for each; a record is covered when some group exists for its own location, which is what the coverage figures below count. Switching status does not vary in BUS39, so every state is a candidate for every query there. In ELES it partitions the population, and the consequences run through the coverage and availability results.

\subsection{Coverage and abstention}\label{sec:res-coverage}

R2 can be met only where the archive holds compatible states, so the first measurement is how often a query finds them. On BUS39, 1,044,793 of 1,044,797 records (99.9996\%) receive a location-specific estimate. All 21,783 states receive at least one estimate, and 21,779 (99.98\%) receive estimates for every recorded contingency. The four missing records have no candidate generator for the required location in the retrieved pool. On ELES, 237,168 of 322,823 records (73.47\%) receive a location-specific estimate. Estimates are available for at least one contingency in 3,344 of 4,393 states (76.12\%), while 2,622 states (59.69\%) have complete coverage.

The ELES shortfall follows directly from the archive partitioning. Its 1,785 topology groups have a median size of one, so most configurations cannot support retrieval. States are distributed unevenly: the median state belongs to a group of four, and coverage is therefore concentrated in the minority of groups that recur. Coverage provides no information about whether an uncovered contingency is safe. A record counts as covered when a candidate generator exists at its recorded location, a weaker requirement than the inclusion of the recorded critical generator among the candidates (Table~\ref{tab:deoracling}).

\subsection{Chronological availability of topology support}\label{sec:res-chronology}

The evaluations above hold out whole operating states but do not impose temporal ordering: an ELES query may retrieve states recorded after it. This protocol is appropriate for measuring retrospective accuracy because a later neighbor does not disclose the query state's target outcome. However, such neighbors would not have been available when the query arrived. To establish what exact-topology retrieval could provide under time-ordered availability, we audit whether each state has an earlier same-topology predecessor without running retrieval (Methods). This state-level availability is necessary but not sufficient because the required location and generator outcome must also be represented in that state.

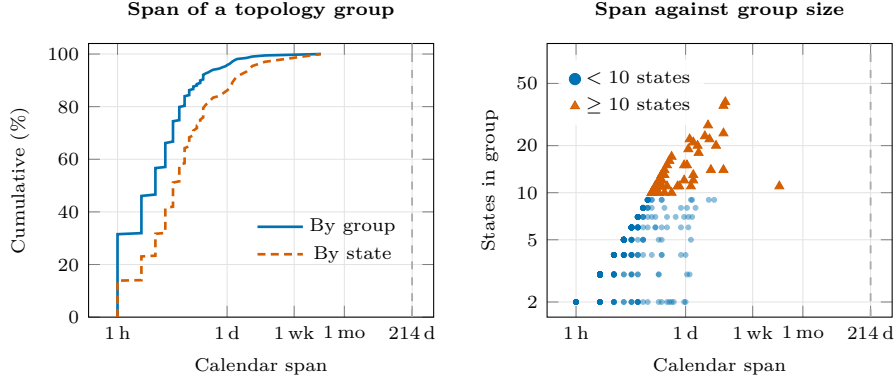
\begin{figure}[tbp]
\centering
\definecolor{okabeBlue}{HTML}{0072B2}
\definecolor{okabeVermillion}{HTML}{D55E00}
\begin{tikzpicture}
\begin{groupplot}[
  group style={group size=2 by 1, horizontal sep=1.5cm},
  width=0.47\textwidth, height=5.2cm,
  xmode=log, xmin=0.018, xmax=400,
  xtick={0.0417,1,7,30,214}, xticklabels={1\,h,1\,d,1\,wk,1\,mo,214\,d},
  tick label style={font=\footnotesize}, label style={font=\footnotesize},
  title style={font=\footnotesize\bfseries},
  grid=major, grid style={gray!22, line width=0.3pt},
  legend style={draw=none, fill=none, font=\footnotesize},
]
\nextgroupplot[title={Span of a topology group}, xlabel={Calendar span}, ylabel={Cumulative (\%)}, ymin=0, ymax=104, legend style={at={(0.93,0.28)}, anchor=east}]
\addplot[gray!65, dashed, line width=0.7pt, mark=none, forget plot] coordinates {(214,0) (214,104)};
\addplot[okabeBlue, line width=1pt, mark=none] coordinates {(0.041667,0.14) (0.041667,0.54) (0.041667,0.95) (0.041667,1.49) (0.041667,1.90) (0.041667,2.31) (0.041667,2.85) (0.041667,3.26) (0.041667,3.67) (0.041667,4.21) (0.041667,4.62) (0.041667,5.03) (0.041667,5.57) (0.041667,5.98) (0.041667,6.39) (0.041667,6.93) (0.041667,7.34) (0.041667,7.88) (0.041667,8.29) (0.041667,8.70) (0.041667,9.24) (0.041667,9.65) (0.041667,10.05) (0.041667,10.60) (0.041667,11.01) (0.041667,11.41) (0.041667,11.96) (0.041667,12.36) (0.041667,12.77) (0.041667,13.32) (0.041667,13.72) (0.041667,14.27) (0.041667,14.67) (0.041667,15.08) (0.041667,15.62) (0.041667,16.03) (0.041667,16.44) (0.041667,16.98) (0.041667,17.39) (0.041667,17.80) (0.041667,18.34) (0.041667,18.75) (0.041667,19.16) (0.041667,19.70) (0.041667,20.11) (0.041667,20.65) (0.041667,21.06) (0.041667,21.47) (0.041667,22.01) (0.041667,22.42) (0.041667,22.83) (0.041667,23.37) (0.041667,23.78) (0.041667,24.18) (0.041667,24.73) (0.041667,25.14) (0.041667,25.54) (0.041667,26.09) (0.041667,26.49) (0.041667,27.04) (0.041667,27.45) (0.041667,27.85) (0.041667,28.40) (0.041667,28.80) (0.041667,29.21) (0.041667,29.76) (0.041667,30.16) (0.041667,30.57) (0.041667,31.11) (0.041667,31.52) (0.083333,31.93) (0.083333,32.47) (0.083333,32.88) (0.083333,33.42) (0.083333,33.83) (0.083333,34.24) (0.083333,34.78) (0.083333,35.19) (0.083333,35.60) (0.083333,36.14) (0.083333,36.55) (0.083333,36.96) (0.083333,37.50) (0.083333,37.91) (0.083333,38.32) (0.083333,38.86) (0.083333,39.27) (0.083333,39.67) (0.083333,40.22) (0.083333,40.62) (0.083333,41.17) (0.083333,41.58) (0.083333,41.98) (0.083333,42.53) (0.083333,42.93) (0.083333,43.34) (0.083333,43.89) (0.083333,44.29) (0.083333,44.70) (0.083333,45.24) (0.083333,45.65) (0.083333,46.06) (0.125,46.60) (0.125,47.01) (0.125,47.55) (0.125,47.96) (0.125,48.37) (0.125,48.91) (0.125,49.32) (0.125,49.73) (0.125,50.27) (0.125,50.68) (0.125,51.09) (0.125,51.63) (0.125,52.04) (0.125,52.45) (0.125,52.99) (0.125,53.40) (0.125,53.94) (0.125,54.35) (0.125,54.76) (0.125,55.30) (0.125,55.71) (0.125,56.11) (0.125,56.66) (0.16667,57.07) (0.16667,57.47) (0.16667,58.02) (0.16667,58.42) (0.16667,58.83) (0.16667,59.38) (0.16667,59.78) (0.16667,60.33) (0.16667,60.73) (0.16667,61.14) (0.16667,61.68) (0.16667,62.09) (0.16667,62.50) (0.16667,63.04) (0.16667,63.45) (0.16667,63.86) (0.16667,64.40) (0.16667,64.81) (0.16667,65.22) (0.16667,65.76) (0.16667,66.17) (0.20833,66.71) (0.20833,67.12) (0.20833,67.53) (0.20833,68.07) (0.20833,68.48) (0.20833,68.89) (0.20833,69.43) (0.20833,69.84) (0.20833,70.24) (0.20833,70.79) (0.20833,71.20) (0.20833,71.60) (0.20833,72.15) (0.20833,72.55) (0.20833,72.96) (0.20833,73.51) (0.20833,73.91) (0.20833,74.46) (0.25,74.86) (0.25,75.27) (0.25,75.82) (0.25,76.22) (0.25,76.63) (0.25,77.17) (0.25,77.58) (0.25,77.99) (0.25,78.53) (0.25,78.94) (0.25,79.35) (0.25,79.89) (0.29167,80.30) (0.29167,80.84) (0.29167,81.25) (0.29167,81.66) (0.29167,82.20) (0.29167,82.61) (0.29167,83.02) (0.29167,83.56) (0.29167,83.97) (0.33333,84.38) (0.33333,84.92) (0.33333,85.33) (0.33333,85.73) (0.33333,86.28) (0.375,86.68) (0.375,87.23) (0.375,87.64) (0.41667,88.04) (0.41667,88.59) (0.45833,88.99) (0.45833,89.40) (0.45833,89.95) (0.5,90.35) (0.5,90.76) (0.5,91.30) (0.5,91.71) (0.5,92.12) (0.54167,92.66) (0.58333,93.07) (0.625,93.61) (0.66667,94.02) (0.79167,94.43) (0.875,94.97) (0.95833,95.38) (1,95.79) (1.0833,96.33) (1.125,96.74) (1.1667,97.15) (1.25,97.69) (1.3333,98.10) (1.75,98.51) (2.0833,99.05) (3,99.46) (15.167,100.00)};
\addplot[okabeVermillion, line width=1pt, densely dashed, mark=none] coordinates {(0.041667,0.06) (0.041667,0.24) (0.041667,0.42) (0.041667,0.66) (0.041667,0.84) (0.041667,1.02) (0.041667,1.26) (0.041667,1.44) (0.041667,1.61) (0.041667,1.85) (0.041667,2.03) (0.041667,2.21) (0.041667,2.45) (0.041667,2.63) (0.041667,2.81) (0.041667,3.05) (0.041667,3.23) (0.041667,3.47) (0.041667,3.65) (0.041667,3.83) (0.041667,4.07) (0.041667,4.25) (0.041667,4.43) (0.041667,4.67) (0.041667,4.84) (0.041667,5.02) (0.041667,5.26) (0.041667,5.44) (0.041667,5.62) (0.041667,5.86) (0.041667,6.04) (0.041667,6.28) (0.041667,6.46) (0.041667,6.64) (0.041667,6.88) (0.041667,7.06) (0.041667,7.24) (0.041667,7.48) (0.041667,7.66) (0.041667,7.83) (0.041667,8.07) (0.041667,8.25) (0.041667,8.43) (0.041667,8.67) (0.041667,8.85) (0.041667,9.09) (0.041667,9.27) (0.041667,9.45) (0.041667,9.69) (0.041667,9.87) (0.041667,10.05) (0.041667,10.29) (0.041667,10.47) (0.041667,10.65) (0.041667,10.89) (0.041667,11.06) (0.041667,11.24) (0.041667,11.48) (0.041667,11.66) (0.041667,11.90) (0.041667,12.08) (0.041667,12.26) (0.041667,12.50) (0.041667,12.68) (0.041667,12.86) (0.041667,13.10) (0.041667,13.28) (0.041667,13.46) (0.041667,13.70) (0.041667,13.88) (0.083333,14.06) (0.083333,14.29) (0.083333,14.50) (0.083333,14.86) (0.083333,15.13) (0.083333,15.40) (0.083333,15.76) (0.083333,16.03) (0.083333,16.30) (0.083333,16.66) (0.083333,16.93) (0.083333,17.14) (0.083333,17.49) (0.083333,17.76) (0.083333,18.03) (0.083333,18.36) (0.083333,18.63) (0.083333,18.90) (0.083333,19.26) (0.083333,19.53) (0.083333,19.89) (0.083333,20.16) (0.083333,20.42) (0.083333,20.78) (0.083333,21.05) (0.083333,21.32) (0.083333,21.68) (0.083333,21.95) (0.083333,22.22) (0.083333,22.58) (0.083333,22.85) (0.083333,23.06) (0.125,23.36) (0.125,23.71) (0.125,24.19) (0.125,24.55) (0.125,24.91) (0.125,25.33) (0.125,25.69) (0.125,26.05) (0.125,26.53) (0.125,26.85) (0.125,27.21) (0.125,27.69) (0.125,27.99) (0.125,28.29) (0.125,28.68) (0.125,29.04) (0.125,29.52) (0.125,29.87) (0.125,30.14) (0.125,30.62) (0.125,30.95) (0.125,31.25) (0.125,31.67) (0.16667,32.00) (0.16667,32.39) (0.16667,32.95) (0.16667,33.37) (0.16667,33.79) (0.16667,34.24) (0.16667,34.69) (0.16667,35.29) (0.16667,35.74) (0.16667,36.18) (0.16667,36.78) (0.16667,37.23) (0.16667,37.62) (0.16667,38.19) (0.16667,38.55) (0.16667,38.91) (0.16667,39.47) (0.16667,39.92) (0.16667,40.34) (0.16667,40.94) (0.16667,41.39) (0.20833,41.99) (0.20833,42.52) (0.20833,43.06) (0.20833,43.60) (0.20833,44.14) (0.20833,44.56) (0.20833,44.98) (0.20833,45.42) (0.20833,45.96) (0.20833,46.65) (0.20833,47.19) (0.20833,47.61) (0.20833,48.21) (0.20833,48.71) (0.20833,49.19) (0.20833,49.88) (0.20833,50.42) (0.20833,51.14) (0.25,51.61) (0.25,52.00) (0.25,52.66) (0.25,53.17) (0.25,53.47) (0.25,54.25) (0.25,54.87) (0.25,55.50) (0.25,56.34) (0.25,56.70) (0.25,57.30) (0.25,57.83) (0.29167,58.49) (0.29167,59.12) (0.29167,59.84) (0.29167,60.38) (0.29167,61.21) (0.29167,61.93) (0.29167,62.65) (0.29167,63.61) (0.29167,64.23) (0.33333,64.95) (0.33333,65.97) (0.33333,66.69) (0.33333,67.49) (0.33333,68.33) (0.375,69.11) (0.375,70.07) (0.375,70.87) (0.41667,71.80) (0.41667,72.79) (0.45833,73.39) (0.45833,74.16) (0.45833,75.03) (0.5,75.90) (0.5,76.35) (0.5,77.48) (0.5,78.59) (0.5,79.46) (0.54167,80.80) (0.58333,81.49) (0.625,82.57) (0.66667,83.43) (0.79167,84.03) (0.875,84.90) (0.95833,85.77) (1,86.15) (1.0833,87.11) (1.125,87.98) (1.1667,89.23) (1.25,90.49) (1.3333,91.48) (1.75,93.30) (2.0833,95.45) (3,97.04) (15.167,100.00)};
\legend{{By group},{By state}}
\nextgroupplot[title={Span against group size}, xlabel={Calendar span}, ylabel={States in group}, ymode=log, ymin=1.6, ymax=90, ytick={2,5,10,20,50}, yticklabels={2,5,10,20,50}, legend image code/.code={\draw[#1] plot coordinates {(0.15cm,0cm)};}, legend style={at={(0.04,0.94)}, anchor=north west, fill=white, fill opacity=0.92, text opacity=1, draw=none}]
\addplot[gray!65, dashed, line width=0.7pt, mark=none, forget plot] coordinates {(214,1.6) (214,90)};
\addplot[only marks, mark=*, mark size=0.9pt, color=okabeBlue, opacity=0.45, legend image post style={mark size=2.1pt, opacity=1}] coordinates {(0.33333,9) (0.33333,9) (0.33333,9) (0.33333,9) (0.33333,9) (1.9583,9) (0.5,9) (2.2917,9) (0.375,9) (0.5,9) (0.33333,9) (0.33333,9) (0.375,9) (0.33333,9) (0.33333,9) (0.33333,9) (0.33333,9) (1.2083,9) (0.33333,9) (1.1667,9) (1.0417,8) (0.29167,8) (0.33333,8) (0.29167,8) (0.29167,8) (0.29167,8) (0.33333,8) (0.29167,8) (0.29167,8) (0.875,8) (0.29167,8) (0.375,8) (1.3333,8) (0.29167,8) (0.29167,8) (0.29167,8) (0.29167,8) (0.29167,8) (0.54167,8) (0.33333,8) (0.29167,8) (0.29167,8) (0.29167,8) (0.29167,8) (0.29167,8) (0.29167,8) (0.29167,8) (0.29167,8) (0.29167,7) (0.41667,7) (1.2083,7) (0.25,7) (0.25,7) (0.75,7) (0.25,7) (0.33333,7) (0.95833,7) (0.25,7) (0.58333,7) (0.25,7) (1.0833,7) (0.25,7) (0.25,7) (0.25,7) (0.25,7) (0.25,7) (0.25,7) (0.25,7) (0.375,7) (0.25,7) (0.29167,7) (0.25,7) (0.25,7) (0.25,7) (0.25,7) (0.25,7) (0.25,7) (0.25,7) (0.29167,7) (0.25,7) (0.875,6) (0.20833,6) (0.20833,6) (0.20833,6) (1.0833,6) (0.20833,6) (0.25,6) (0.20833,6) (0.5,6) (0.25,6) (0.20833,6) (0.20833,6) (0.20833,6) (0.20833,6) (0.25,6) (0.20833,6) (0.20833,6) (0.20833,6) (0.25,6) (0.25,6) (0.20833,6) (0.20833,6) (0.20833,6) (0.33333,6) (0.20833,6) (0.20833,6) (0.20833,6) (0.20833,6) (0.20833,6) (0.20833,6) (0.20833,6) (0.58333,6) (0.20833,6) (0.20833,6) (0.20833,6) (0.25,6) (0.20833,6) (0.20833,6) (0.20833,6) (0.20833,6) (0.41667,6) (0.41667,6) (0.29167,6) (0.20833,6) (0.20833,6) (0.20833,6) (0.20833,6) (0.20833,6) (0.20833,6) (0.20833,6) (0.20833,6) (0.25,6) (0.20833,6) (0.20833,6) (0.20833,6) (0.20833,6) (0.20833,6) (0.25,6) (0.20833,5) (0.16667,5) (0.16667,5) (0.16667,5) (0.20833,5) (0.20833,5) (0.16667,5) (0.16667,5) (0.25,5) (0.16667,5) (0.20833,5) (0.45833,5) (0.16667,5) (0.16667,5) (0.16667,5) (0.16667,5) (0.16667,5) (0.16667,5) (0.16667,5) (0.16667,5) (0.16667,5) (0.16667,5) (0.16667,5) (0.16667,5) (0.16667,5) (0.16667,5) (0.16667,5) (0.16667,5) (0.16667,5) (0.16667,5) (0.16667,5) (0.16667,5) (0.16667,5) (0.16667,5) (0.29167,5) (0.16667,5) (0.16667,5) (0.16667,5) (0.16667,5) (0.5,5) (0.16667,5) (0.16667,5) (0.16667,5) (0.16667,5) (0.16667,5) (0.29167,5) (0.16667,5) (0.16667,5) (0.16667,5) (0.16667,5) (0.16667,5) (0.16667,5) (0.16667,5) (0.16667,5) (0.16667,5) (0.16667,5) (0.20833,5) (0.16667,5) (0.16667,5) (0.29167,5) (0.16667,5) (0.16667,5) (0.16667,5) (0.16667,5) (0.16667,5) (0.16667,5) (0.20833,5) (1.1667,5) (0.16667,5) (0.125,4) (0.125,4) (0.16667,4) (0.16667,4) (0.125,4) (0.125,4) (0.125,4) (0.125,4) (0.125,4) (1,4) (0.125,4) (0.20833,4) (0.16667,4) (0.20833,4) (0.125,4) (0.125,4) (0.125,4) (0.16667,4) (0.125,4) (0.16667,4) (0.125,4) (0.125,4) (0.125,4) (0.125,4) (0.125,4) (0.20833,4) (0.375,4) (0.125,4) (0.125,4) (0.125,4) (0.125,4) (0.125,4) (0.20833,4) (0.125,4) (0.16667,4) (0.875,4) (0.125,4) (0.125,4) (0.5,4) (0.125,4) (0.20833,4) (0.125,4) (0.20833,4) (0.125,4) (0.16667,4) (0.125,4) (0.125,4) (0.25,4) (0.125,4) (0.125,4) (0.125,4) (0.125,4) (0.125,4) (0.125,4) (0.16667,4) (0.25,4) (0.125,4) (0.125,4) (0.125,4) (0.125,4) (0.125,4) (0.29167,4) (0.125,4) (0.125,4) (0.125,4) (0.125,4) (0.125,4) (0.125,4) (0.125,4) (0.125,4) (0.25,4) (0.125,4) (0.125,4) (0.125,4) (0.16667,4) (0.16667,4) (0.16667,4) (0.5,4) (0.125,4) (0.125,4) (0.125,4) (0.125,4) (0.125,4) (0.125,4) (0.125,4) (0.125,4) (0.125,4) (0.083333,3) (0.083333,3) (0.083333,3) (0.083333,3) (0.083333,3) (0.083333,3) (0.083333,3) (0.083333,3) (0.125,3) (0.20833,3) (0.125,3) (0.16667,3) (0.083333,3) (0.125,3) (0.083333,3) (0.083333,3) (0.083333,3) (0.083333,3) (0.083333,3) (0.083333,3) (0.125,3) (0.083333,3) (0.083333,3) (0.083333,3) (0.20833,3) (0.083333,3) (0.083333,3) (0.083333,3) (0.083333,3) (0.083333,3) (0.083333,3) (0.083333,3) (0.083333,3) (0.083333,3) (0.20833,3) (0.083333,3) (0.083333,3) (0.083333,3) (0.083333,3) (0.125,3) (0.083333,3) (0.083333,3) (0.16667,3) (0.083333,3) (0.083333,3) (0.083333,3) (0.083333,3) (0.083333,3) (0.083333,3) (0.083333,3) (0.083333,3) (0.083333,3) (0.083333,3) (0.083333,3) (0.083333,3) (0.083333,3) (0.083333,3) (0.083333,3) (0.083333,3) (0.25,3) (0.125,3) (0.083333,3) (0.125,3) (0.083333,3) (0.083333,3) (0.083333,3) (0.083333,3) (0.45833,3) (0.083333,3) (0.083333,3) (0.083333,3) (0.083333,3) (0.083333,3) (0.125,3) (0.083333,3) (0.125,3) (0.083333,3) (0.083333,3) (0.083333,3) (0.083333,3) (0.083333,3) (0.083333,3) (0.125,3) (0.083333,3) (0.083333,3) (0.083333,3) (0.083333,3) (0.083333,3) (0.083333,3) (0.083333,3) (0.083333,3) (0.083333,3) (0.083333,3) (0.083333,3) (0.083333,3) (0.125,3) (0.25,3) (0.083333,3) (0.083333,3) (0.083333,3) (0.083333,3) (0.083333,3) (0.083333,3) (0.083333,3) (0.083333,3) (1.0417,3) (0.083333,3) (0.125,3) (0.083333,3) (1.125,3) (0.083333,3) (0.083333,3) (0.45833,3) (0.083333,3) (0.083333,3) (0.083333,3) (0.041667,2) (0.041667,2) (0.041667,2) (0.041667,2) (0.041667,2) (0.041667,2) (0.041667,2) (0.041667,2) (0.041667,2) (0.041667,2) (0.041667,2) (0.083333,2) (0.041667,2) (0.041667,2) (0.083333,2) (0.041667,2) (0.041667,2) (0.041667,2) (0.041667,2) (0.041667,2) (0.041667,2) (0.041667,2) (0.041667,2) (0.041667,2) (0.125,2) (0.041667,2) (0.041667,2) (0.041667,2) (0.083333,2) (0.041667,2) (0.20833,2) (0.041667,2) (0.041667,2) (0.041667,2) (0.041667,2) (0.083333,2) (0.041667,2) (0.041667,2) (0.041667,2) (0.041667,2) (0.083333,2) (0.083333,2) (0.041667,2) (0.041667,2) (0.041667,2) (0.041667,2) (0.041667,2) (0.041667,2) (0.041667,2) (0.041667,2) (0.041667,2) (0.041667,2) (0.041667,2) (0.041667,2) (0.041667,2) (0.041667,2) (0.041667,2) (0.041667,2) (0.041667,2) (0.041667,2) (0.041667,2) (0.041667,2) (0.25,2) (0.041667,2) (0.041667,2) (0.041667,2) (0.041667,2) (0.25,2) (0.041667,2) (0.041667,2) (0.041667,2) (0.041667,2) (0.041667,2) (0.083333,2) (0.041667,2) (0.041667,2) (0.041667,2) (0.041667,2) (0.041667,2) (0.041667,2) (0.041667,2) (0.041667,2) (0.041667,2) (0.041667,2) (0.041667,2) (0.5,2) (0.041667,2) (0.041667,2) (0.041667,2) (0.041667,2) (0.041667,2) (0.041667,2) (0.041667,2) (0.041667,2) (0.041667,2) (0.041667,2) (0.041667,2) (0.29167,2) (0.20833,2) (0.041667,2) (0.041667,2) (0.041667,2) (0.041667,2) (0.041667,2) (0.041667,2) (0.041667,2) (0.041667,2) (0.25,2) (0.041667,2) (0.041667,2) (0.041667,2) (0.041667,2) (0.041667,2) (0.041667,2) (0.041667,2) (0.66667,2) (0.041667,2) (0.041667,2) (0.041667,2) (0.041667,2) (0.041667,2) (0.041667,2) (0.041667,2) (0.041667,2) (0.041667,2) (0.083333,2) (0.125,2) (0.041667,2) (0.041667,2) (0.041667,2) (0.041667,2) (0.125,2) (0.083333,2) (0.041667,2) (0.041667,2) (0.041667,2) (0.041667,2) (0.041667,2) (0.041667,2) (0.083333,2) (0.041667,2) (0.041667,2) (0.041667,2) (0.041667,2) (0.041667,2) (0.041667,2) (0.041667,2) (0.041667,2) (0.041667,2) (0.041667,2) (0.041667,2) (0.041667,2) (0.041667,2) (0.041667,2) (0.041667,2) (0.041667,2) (0.041667,2) (0.95833,2) (0.041667,2) (0.041667,2) (1,2) (0.20833,2) (0.041667,2) (0.041667,2) (0.041667,2) (0.58333,2) (0.041667,2) (0.041667,2) (0.041667,2) (0.041667,2) (0.041667,2) (0.041667,2) (0.041667,2) (0.041667,2) (0.25,2) (0.041667,2) (0.041667,2) (0.041667,2) (0.041667,2) (0.041667,2) (0.041667,2) (0.041667,2) (0.25,2) (0.041667,2) (0.041667,2) (0.041667,2) (0.041667,2) (0.041667,2) (0.041667,2) (0.041667,2) (0.041667,2) (0.041667,2) (0.041667,2) (0.041667,2) (0.041667,2) (0.041667,2) (0.041667,2) (0.16667,2) (0.125,2) (0.041667,2) (0.041667,2) (0.041667,2) (0.041667,2) (0.041667,2) (0.041667,2) (0.041667,2) (0.041667,2) (0.70833,2) (0.041667,2) (0.041667,2) (0.041667,2) (0.041667,2) (0.083333,2) (0.041667,2) (0.041667,2) (0.041667,2) (0.083333,2) (0.041667,2) (0.041667,2) (0.041667,2) (0.041667,2) (0.16667,2) (0.041667,2) (0.041667,2) (0.45833,2) (0.041667,2) (0.041667,2) (0.041667,2) (0.041667,2) (0.083333,2) (0.041667,2) (0.041667,2) (0.041667,2) (0.041667,2) (0.625,2) (0.041667,2) (0.041667,2) (0.041667,2) (0.041667,2) (0.041667,2) (0.041667,2) (0.041667,2) (0.041667,2) (0.041667,2) (0.041667,2) (0.041667,2) (0.041667,2) (0.041667,2) (0.041667,2) (0.041667,2) (0.041667,2) (0.041667,2) (0.041667,2) (0.041667,2) (0.041667,2) (0.041667,2) (0.041667,2) (0.041667,2) (0.041667,2) (0.041667,2) (0.041667,2) (0.041667,2) (0.041667,2) (0.083333,2) (0.041667,2) (0.125,2) (0.041667,2) (0.041667,2) (0.041667,2) (0.041667,2) (0.041667,2) (0.33333,2) (0.041667,2)};
\addplot[only marks, mark=triangle*, mark size=1.9pt, color=okabeVermillion] coordinates {(3.1667,38) (3,36) (1.9167,27) (3,24) (1.75,23) (1.125,22) (2,22) (1.25,21) (2.4167,20) (1.4167,20) (1.0833,19) (1.4583,18) (0.66667,17) (0.625,16) (0.95833,15) (1.0417,15) (0.58333,15) (3,14) (2.0833,14) (0.54167,14) (0.5,13) (0.5,13) (0.54167,13) (1.25,13) (0.5,13) (0.5,13) (0.95833,12) (0.5,12) (0.5,12) (0.45833,12) (1.25,12) (0.5,11) (1.1667,11) (0.83333,11) (0.79167,11) (0.41667,11) (0.5,11) (0.41667,11) (15.167,11) (0.45833,11) (0.41667,11) (0.45833,11) (0.58333,11) (0.5,11) (0.375,10) (0.375,10) (0.375,10) (0.375,10) (0.375,10) (0.54167,10) (0.45833,10) (0.41667,10) (0.66667,10)};
\legend{{$<10$ states},{$\geq10$ states}}
\end{groupplot}
\end{tikzpicture}
\caption{Calendar distribution of ELES topology groups under the deployed line-status key, over a 214-day record. Left: cumulative distribution of a group's calendar span, the time between its first and last state, weighting the 736 multi-state groups equally and then weighting each state equally, since large groups hold most states. Right: the same span against group size, which tests whether the largest groups are the durable ones; they are not. The dashed line marks the record length in both panels. Groups holding one state are excluded, as a span is undefined for them; there are 1,049 such groups.}
\label{fig:burstiness}
\end{figure}

Under continuous ingestion, where any strictly earlier state is available, 59.37\% of ELES states have an earlier same-topology candidate, against 76.12\% receiving at least one estimate retrospectively. Under a reference library frozen at a cutoff, support for subsequent states falls to between 0 and 2.84\% across nine pre-specified calendar deciles. The reason is that ELES topology groups are temporally concentrated (Fig.~\ref{fig:burstiness}). Of the 1,785 line-status groups, 1,049 hold a single state; the rest span a median of three hours in a 214-day record, and 99.3\% fall in one cluster at a 24-hour split, so a configuration rarely returns after its burst. The 53 groups with ten or more states span a median of 14 hours. A continuously updated library can therefore draw on earlier states from the same burst, while a frozen one cannot. Delaying every state's availability by up to an hour changes no figure here, since states are sampled hourly and no same-topology pair is less than 60 minutes apart.

These are two scenarios, not bounds on every deployment: continuous ingestion assumes zero delay, and the frozen library assumes no updates. Neither is a prospective evaluation of estimation accuracy.

\subsection{Retrieval under causal availability}\label{sec:res-nearduplicate}

The audit above counts candidates; this subsection runs retrieval under the same restriction. A deployment may use any state recorded before its query whose contingencies have finished simulating, and none recorded after it. We evaluated ELES under that rule and repeated it with an hour of ingestion delay (Methods). Because states are sampled hourly the two are identical: any delay up to the sampling interval removes no candidate.

Time ordering substantially reduces coverage while producing a more modest increase in conditional error. Of 322,823 records, 181,868 (56.34\%) receive an estimate, compared with 73.47\% retrospectively, and every causally covered record is also covered retrospectively. On these 181,868 records, paired \acs{mae} increases from \qty{0.0482}{\second} to \qty{0.0530}{\second}, an increase of 10.0\% (95\% CI 0.0038 to 0.0057\,s; paired Wilcoxon signed-rank test on the mean absolute errors of 2,608 states, $W=560{,}591.5$, $p=7.3\times10^{-35}$, $r=0.288$), whereas the median absolute error remains \qty{0.010}{\second}. Critical-generator recovery decreases from 70.82\% retrospectively to 69.54\% causally. Time ordering therefore primarily reduces coverage rather than critical-generator identification. Under the adopted rule, coverage loss is irreducible because every unanswered contingency requires referral, whereas increased prediction error affects referral only when an estimate crosses the threshold. Loss of compatible coverage is therefore the principal deployment constraint in this archive.

Withholding states on both sides of the query provides a conservative assessment of retrieval's dependence on temporal neighbors. This is a stress test rather than a deployment model because it removes recent history that would legitimately be available in deployment. Excluding every state within three hours increases error by 76.7\%, and widening the window to 48 hours produces little additional increase. Each window scores a different surviving population against its own baseline, so these are separate paired comparisons rather than one series: the large increase is already present at three hours in every one of them, but their magnitudes are not directly comparable across windows. For a gradient-boosting regressor refitted under the same temporal-exclusion condition, error increases by 8.6\%, approximately one-ninth of the retrieval increase. Retrieval accuracy in this archive therefore depends strongly on states recorded within a few hours of the query, which qualifies the comparison with supervised models below (Supplementary Note~S1, Supplementary Fig.~S1).

\subsection{Operational screening at clearing-time thresholds}\label{sec:res-screening}

\begin{table}[htbp]
\centering
\caption{Minimum-estimate screening over the evaluated contingencies. A state is referred if its minimum estimate is below $\tau$ or any recorded contingency lacks an estimate. Bootstrap 95\% CIs use 2,000 state-level resamples (Methods). Lift is recall minus state referral: referring a uniformly random share of states yields an expected recall equal to that share, so lift is the recall gained over no-skill referral at the same workload. Rows differ in $n_\mathrm{unsafe}$ and are therefore separate classification problems, so lift is comparable within a row rather than across thresholds. Where $n_\mathrm{unsafe}=0$ both recall and lift are undefined, and every referral is an abstention, so the false-alarm rate equals the state-referral rate by construction and measures coverage rather than discrimination; that row is not comparable with the rows above it.}
\label{tab:screening}
\footnotesize
\setlength{\tabcolsep}{2pt}
\begin{tabular}{lrrrrrr}
\toprule
& $\tau$ (ms) & Recall (95\% CI) & False-alarm (95\% CI) & State referral (95\% CI) & Lift (pts) & $n_\mathrm{unsafe}$ \\
\midrule
\multirow{3}{*}{BUS39} & 100 & 70.23\% (66.40, 73.79) & 0.14\% (0.09, 0.19) & 1.79\% (1.62, 1.96) & 68.4 & 514 \\
& 150 & 90.85\% (89.63, 92.00) & 1.43\% (1.26, 1.59) & 10.49\% (10.08, 10.89) & 80.4 & 2,207 \\
& 200 & 94.47\% (93.86, 95.06) & 2.56\% (2.33, 2.81) & 27.49\% (26.90, 28.11) & 67.0 & 5,908 \\
\midrule
\multirow{3}{*}{ELES} & 100 & undefined & 40.31\% (38.86, 41.77) & 40.31\% (38.86, 41.77) & undefined & 0 \\
& 150 & 92.49\% (90.27, 94.36) & 43.71\% (42.14, 45.23) & 51.10\% (49.65, 52.54) & 41.4 & 666 \\
& 200 & 97.52\% (96.93, 98.10) & 54.33\% (51.90, 56.74) & 81.27\% (80.08, 82.40) & 16.3 & 2,740 \\
\bottomrule
\end{tabular}
\end{table}

The coverage shortfall produced by abstention translates into operational workload because each abstention requires referral; recall must therefore be interpreted jointly with referral load. At $\tau=\qty{200}{\ms}$, scored retrospectively, BUS39 achieves 94.47\% recall while referring 27.49\% of states, and ELES achieves 97.52\% recall while referring 81.27\%. Lift over no-skill referral is 67.0 percentage points on BUS39 and 16.3 on ELES, but it is bounded by referral load and is meaningfully comparable within a row rather than across thresholds. Much of the ELES referral load is unavoidable: 62.37\% of its states are observed unsafe at this threshold, so any screen with full recall must refer at least this proportion. Of the remaining 37.63 percentage points of potentially avoidable simulations, retrieval avoids 18.73 points. The unsafe proportion is a property of the threshold and evaluated contingency set, not an assessment of the system as operated. The remaining referral load is also not specific to retrieval. On the five-fold population, with each comparator's cutoff fixed on the folds to which a state does not belong, gradient boosting at retrieval's referral load achieves 96.42\% recall, compared with 96.82\% for retrieval. The 0.40-percentage-point difference has an interval spanning zero (95\% CI $-0.74$ to $+1.60$, topology groups resampled). Retrieval cannot operate below a referral rate of 42.70\% on that population because this proportion of states contains a contingency without compatible support. Archive coverage therefore establishes a minimum referral workload, to which threshold crossings contribute. No ELES state is observed unsafe at 100\,ms under the evaluated contingencies, leaving recall undefined.

\subsection{Critical-generator selection and point accuracy}\label{sec:res-deoracling}

The results above quantify the consequences of R2. R1 is enforced by construction: retrieval retains the supporting cases, their outcome grouping, and the explicit aggregation rule. This representation enables the analysis below, which evaluates selected estimates where support exists. For each evaluated contingency, we hold the recorded fault location fixed and rank candidate critical generators by unnormalized kernel-support mass (Methods). The oracle row uses the estimate for the recorded critical generator, whereas the selected-generator row uses the candidate with the greatest support. The experiment evaluates critical-generator identification at a known fault location, which is the conditional form of the task considered here.

\begin{table}[htbp]
\centering
\caption{CCT \acl{mae} (\acs{mae}) at the recorded fault location. Bootstrap 95\% CIs use 1,000 state-level resamples (Methods). Each row uses its own covered population, so the rows are not paired: they differ by the records whose recorded critical generator is absent from the retrieved candidates, 405 on BUS39 and 31,735 on ELES. Supplementary Note~S2 gives the paired comparison.}
\label{tab:deoracling}
\footnotesize
\setlength{\tabcolsep}{4pt}
\begin{tabular}{lrrrr}
\toprule
& \multicolumn{2}{c}{BUS39} & \multicolumn{2}{c}{ELES} \\
Level & \acs{mae} (s; 95\% CI) & $n$ & \acs{mae} (s; 95\% CI) & $n$ \\
\midrule
True generator (oracle) & 0.0182 (0.0181, 0.0183) & 1,044,388 & 0.0429 (0.0418, 0.0441) & 205,433 \\
Highest-support generator & 0.0190 (0.0189, 0.0191) & 1,044,793 & 0.0497 (0.0484, 0.0511) & 237,168 \\
\bottomrule
\end{tabular}
\end{table}

Retrieval reaches that estimate without being told the recorded critical generator. The candidate set included the recorded critical generator in 99.96\% of BUS39 cases and 86.62\% of ELES cases, and the highest-support candidate matched it in 95.5\% and 81.8\% of those, giving selected-generator \acs{mae} of \qty{0.0190}{\second} and \qty{0.0497}{\second} (Table~\ref{tab:deoracling}). On ELES the selected generator gives lower error than the recorded one on common support, so the oracle is not an error lower bound there. Supplementary Note~S2 reports that comparison and the population difference behind it.

We compare point accuracy with three supervised regressors, among them extra trees~\cite{geurts2006extratrees}, on records covered by retrieval (Table~\ref{tab:extratrees}). All use the same five state-group folds, pre-fault variables and recorded location; none receives the recorded critical generator, terminal or type. Retrieval selects the highest-support candidate generator. This comparison uses a separate five-fold ELES retrieval run, not the leave-one-state-out population used elsewhere. Removing a whole test fold from the reference library, rather than only the query state, empties more of ELES's small topology groups, so coverage decreases from 73.47 to 71.40\%. On BUS39, the same change excludes one additional record: 1,044,792 records remain covered, compared with 1,044,793.

\begin{table}[htbp]
\centering
\caption{Point error on the records covered by retrieval in both arms, under matched state-group folds. Errors are the \acl{mae} and the \acl{rmse} (\acs{rmse}). Coverage is relative to all records; the location-median baseline is fitted within each training fold. $q_{50}$ to $q_{100}$ are absolute-error quantiles in seconds, with $q_{100}$ the maximum. Values are rounded to three decimals, the resolution the bootstrap intervals support; ratios quoted in the text are computed from unrounded values. Tree ensembles are labeled by their per-split feature budget $p$, which is part of the comparator's identity: both budgets are reported on BUS39 and $\sqrt{p}$ on ELES, where evaluating every feature was intractable (Methods). Gradient boosting has no such budget.}
\label{tab:extratrees}
\footnotesize
\setlength{\tabcolsep}{2pt}
\begin{tabular}{llrrrrrrrr}
\toprule
Dataset & Method & $n$ & Coverage (\%) & \acs{mae} (s) & \acs{rmse} (s) & $q_{50}$ & $q_{95}$ & $q_{99}$ & $q_{100}$ \\
\midrule
\multirow{7}{*}{BUS39} & Retrieval & 1,044,792 & 99.9995 & 0.019 & 0.031 & 0.013 & 0.054 & 0.120 & 0.801 \\
& Extra trees (all $p$) & 1,044,792 & 99.9995 & 0.005 & 0.010 & 0.003 & 0.015 & 0.038 & 0.536 \\
& Random forest (all $p$) & 1,044,792 & 99.9995 & 0.005 & 0.010 & 0.003 & 0.016 & 0.038 & 0.555 \\
& Random forest ($\sqrt{p}$) & 1,044,792 & 99.9995 & 0.021 & 0.030 & 0.018 & 0.050 & 0.098 & 0.770 \\
& Extra trees ($\sqrt{p}$) & 1,044,792 & 99.9995 & 0.019 & 0.028 & 0.015 & 0.047 & 0.095 & 0.724 \\
& Gradient boosting & 1,044,792 & 99.9995 & 0.016 & 0.030 & 0.010 & 0.044 & 0.110 & 0.983 \\
& Location median & 1,044,792 & 99.9995 & 0.111 & 0.161 & 0.080 & 0.300 & 0.590 & 1.710 \\
\multirow{5}{*}{ELES} & Retrieval & 230,511 & 71.40 & 0.051 & 0.120 & 0.010 & 0.230 & 0.590 & 1.620 \\
& Gradient boosting & 230,511 & 71.40 & 0.055 & 0.118 & 0.023 & 0.213 & 0.585 & 1.469 \\
& Random forest ($\sqrt{p}$) & 230,511 & 71.40 & 0.104 & 0.156 & 0.075 & 0.290 & 0.636 & 1.455 \\
& Extra trees ($\sqrt{p}$) & 230,511 & 71.40 & 0.102 & 0.153 & 0.073 & 0.284 & 0.626 & 1.466 \\
& Location median & 230,511 & 71.40 & 0.091 & 0.158 & 0.050 & 0.320 & 0.700 & 1.570 \\
\bottomrule
\end{tabular}
\end{table}

Retrieval performs differently relative to the supervised baselines in the two datasets, and the estimated difference depends strongly on the comparator. On BUS39, retrieval error is 3.9 times that of extra trees at the full feature budget but approximately equal to that of the same model at $\sqrt{p}$, so the difference reflects feature budget as well as method. Retrieval error is 1.2 times that of gradient boosting, although retrieval has the smaller maximum error, \qty{0.801}{\second} compared with \qty{0.983}{\second}. On the matched ELES population, retrieval has the lowest \acs{mae}, 7.7\% below gradient boosting and approximately 50\% below the $\sqrt{p}$ ensembles, but that advantage is confined to the bulk of the error distribution: it leads at the median, 0.0100 against 0.0227\,s, while the two are level at the 99th percentile and gradient boosting has the smaller maximum. Satisfying the requirements therefore involves archive-dependent accuracy trade-offs and does not guarantee improved accuracy.

\subsection{Which retrieval-support diagnostics track error}\label{sec:res-diagnostics}

\begin{figure}[tbp]
\centering
\definecolor{okabeBlue}{HTML}{0072B2}
\definecolor{okabeOrange}{HTML}{E69F00}
\definecolor{okabeGreen}{HTML}{009E73}
\definecolor{okabeVermillion}{HTML}{D55E00}
\definecolor{okabePurple}{HTML}{CC79A7}
\definecolor{okabeSky}{HTML}{56B4E9}
\definecolor{okabeBlack}{HTML}{000000}
\begin{tikzpicture}
\begin{groupplot}[
  group style={group size=2 by 1, horizontal sep=1.5cm},
  width=0.47\textwidth, height=5.4cm,
  xmin=0.47, xmax=1.03, xtick={0.5,0.6,0.7,0.8,0.9,1.0},
  xlabel={Coverage}, ylabel={MAE relative to full coverage},
  tick label style={font=\footnotesize}, label style={font=\footnotesize},
  title style={font=\footnotesize\bfseries},
  grid=major, grid style={gray!22, line width=0.3pt},
  every axis plot/.append style={line width=0.9pt, mark size=1.6pt},
  legend style={draw=none, fill=none, font=\footnotesize, legend columns=3,
                /tikz/every even column/.append style={column sep=10pt}},
]
\nextgroupplot[title={BUS39}, ymin=0.57, ymax=1.13, ytick={0.6,0.7,0.8,0.9,1.0,1.1}, legend style={at={(1.12,-0.26)}, anchor=north}]
\addplot[gray!65, dashed, line width=0.7pt, mark=none, forget plot] coordinates {(0.47,1.0) (1.03,1.0)};
\addplot[color=okabeBlue, mark=*, mark options={fill=okabeBlue, solid}] coordinates {(0.50,0.6684) (0.70,0.7307) (0.80,0.7664) (0.90,0.8179) (0.95,0.8628) (1.00,1.0000)};
\addplot[color=okabeSky, mark=oplus*, mark options={fill=okabeSky, solid}] coordinates {(0.50,0.8745) (0.70,0.8745) (0.80,0.8964) (0.90,0.9476) (0.95,0.9676) (1.00,1.0000)};
\addplot[color=okabeBlack, mark=star, mark options={fill=okabeBlack, solid}, dashed] coordinates {(0.50,0.8745) (0.70,0.8745) (0.80,0.8964) (0.90,0.9476) (0.95,0.9676) (1.00,1.0000)};
\addplot[color=okabeOrange, mark=square*, mark options={fill=okabeOrange, solid}] coordinates {(0.50,0.9015) (0.70,0.8757) (0.80,0.9110) (0.90,0.9321) (0.95,0.9399) (1.00,1.0000)};
\addplot[color=okabeGreen, mark=triangle*, mark options={fill=okabeGreen, solid}] coordinates {(0.50,0.9293) (0.70,0.9436) (0.80,0.9599) (0.90,0.9807) (0.95,0.9913) (1.00,1.0000)};
\addplot[color=okabeVermillion, mark=diamond*, mark options={fill=okabeVermillion, solid}] coordinates {(0.50,1.0613) (0.70,1.0443) (0.80,1.0297) (0.90,1.0135) (0.95,1.0058) (1.00,1.0000)};
\addplot[color=okabePurple, mark=pentagon*, mark options={fill=okabePurple, solid}] coordinates {(0.50,1.0863) (0.70,1.0593) (0.80,1.0419) (0.90,1.0191) (0.95,1.0091) (1.00,1.0000)};
\legend{
  {Weighted CCT dispersion},
  {Retrieved neighbor count},
  {Unique supporting states},
  {Location support mass},
  {Effective sample size},
  {Minimum query distance},
  {Neighborhood compactness}
}
\nextgroupplot[title={ELES}, ymin=0.57, ymax=1.13, ytick={0.6,0.7,0.8,0.9,1.0,1.1}]
\addplot[gray!65, dashed, line width=0.7pt, mark=none, forget plot] coordinates {(0.47,1.0) (1.03,1.0)};
\addplot[color=okabeBlue, mark=*, mark options={fill=okabeBlue, solid}] coordinates {(0.50,0.8413) (0.70,0.8707) (0.80,0.8818) (0.90,0.8906) (0.95,0.9198) (1.00,1.0000)};
\addplot[color=okabeSky, mark=oplus*, mark options={fill=okabeSky, solid}] coordinates {(0.50,0.8640) (0.70,0.9046) (0.80,0.9355) (0.90,0.9713) (0.95,0.9864) (1.00,1.0000)};
\addplot[color=okabeBlack, mark=star, mark options={fill=okabeBlack, solid}, dashed] coordinates {(0.50,0.8419) (0.70,0.8958) (0.80,0.9283) (0.90,0.9536) (0.95,0.9746) (1.00,1.0000)};
\addplot[color=okabeOrange, mark=square*, mark options={fill=okabeOrange, solid}] coordinates {(0.50,1.0024) (0.70,0.9320) (0.80,0.9497) (0.90,0.9839) (0.95,0.9852) (1.00,1.0000)};
\addplot[color=okabeGreen, mark=triangle*, mark options={fill=okabeGreen, solid}] coordinates {(0.50,1.0049) (0.70,0.9723) (0.80,0.9783) (0.90,0.9831) (0.95,0.9927) (1.00,1.0000)};
\addplot[color=okabeVermillion, mark=diamond*, mark options={fill=okabeVermillion, solid}] coordinates {(0.50,0.6645) (0.70,0.7748) (0.80,0.8355) (0.90,0.9059) (0.95,0.9528) (1.00,1.0000)};
\addplot[color=okabePurple, mark=pentagon*, mark options={fill=okabePurple, solid}] coordinates {(0.50,0.9942) (0.70,0.9566) (0.80,0.9568) (0.90,0.9722) (0.95,0.9863) (1.00,1.0000)};
\end{groupplot}
\end{tikzpicture}
\caption{Risk-coverage curves for the seven support diagnostics, on the same populations as Table~\ref{tab:diagnostics}. Records are ordered from most to least trusted by each diagnostic, and \acs{mae} is recomputed as the least-trusted records are withheld. Each curve is divided by its own full-coverage \acs{mae}, the normalization inside \acs{naurc}, which places both datasets on one scale. The dashed line at 1.0 is the unsorted baseline: below it a diagnostic withholds the records it should, above it withholding by that diagnostic is worse than withholding at random. The two count diagnostics coincide on BUS39 by construction, since $K=100$ binds there and each retrieved state contributes one neighbor, so their curves overlap in that panel and the two rows are one result rather than two.}
\label{fig:risk-coverage}
\end{figure}
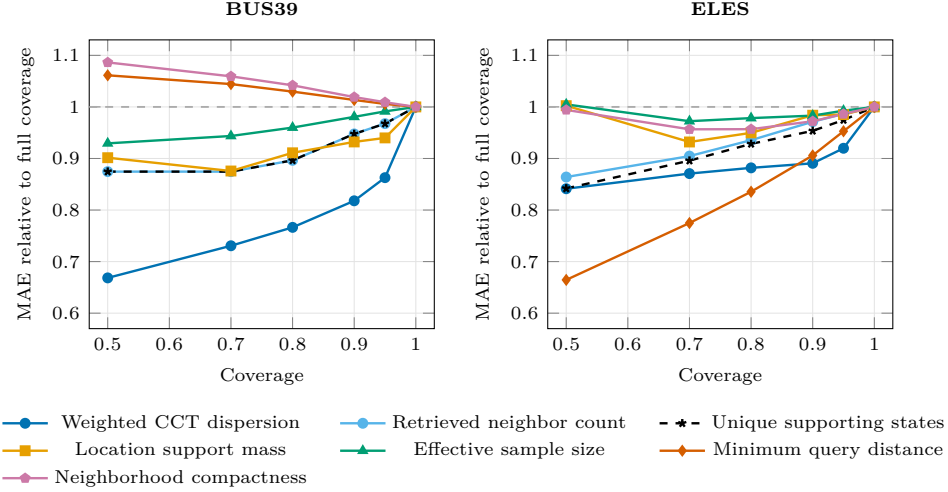

R3 requires support statistics to be evaluated rather than asserted, so we measure whether each diagnostic ranks estimates according to their observed error. No diagnostic ranks first on both datasets (Table~\ref{tab:diagnostics}, Fig.~\ref{fig:risk-coverage}).

\begin{table}[htbp]
\centering
\small
\caption{\Acl{naurc} (\acs{naurc}) against highest-support-generator error. Bootstrap 95\% CIs use 1,000 state-level resamples. Values below 1.0 are favorable; values above 1.0 mean the diagnostic does worse than not sorting at all over the 50 to 100\% coverage range integrated here. Weighted \acs{cct} dispersion and neighborhood compactness require at least two retrieved neighbors and are restricted to the records where both exist; the other five are defined for every covered record. Evaluated on all 237,168 covered ELES records the two count diagnostics give 0.893 (0.882, 0.903) and 0.887 (0.876, 0.897).}
\label{tab:diagnostics}
\begin{tabular}{lrr}
\toprule
Diagnostic & BUS39 \acs{naurc} (95\% CI) & ELES \acs{naurc} (95\% CI) \\
\midrule
Weighted \acs{cct} dispersion & 0.765 (0.763, 0.767) & 0.882 (0.869, 0.893) \\
Retrieved neighbor count & 0.905 (0.903, 0.907) & 0.926 (0.915, 0.936) \\
Unique supporting states & 0.905 (0.903, 0.907) & 0.913 (0.901, 0.925) \\
Location support mass & 0.909 (0.906, 0.912) & 0.966 (0.956, 0.977) \\
Effective sample size & 0.957 (0.954, 0.960) & 0.985 (0.975, 0.995) \\
Minimum query distance & 1.034 (1.031, 1.037) & 0.814 (0.799, 0.828) \\
Neighborhood compactness & 1.047 (1.044, 1.050) & 0.972 (0.962, 0.982) \\
\bottomrule
\end{tabular}
\end{table}

\begin{sloppypar}
Weighted \acs{cct} dispersion, which measures agreement among retrieved outcomes, leads on BUS39 (0.765); on ELES minimum query-to-neighbor distance leads (0.814) with dispersion second (0.882). The same distance diagnostic reverses the desired ordering on BUS39 (1.034, CI excluding 1.0), so its ELES performance does not transfer. The intervals describe each diagnostic separately and do not test differences between them. Effective sample size and neighborhood compactness remain close to the uninformative value of 1.0 on both datasets.

Of the dataset-specific leading diagnostics, only weighted dispersion remains informative in both datasets. The two count diagnostics are also informative in both, at 0.905 on BUS39 and 0.926 and 0.913 on ELES, with intervals excluding 1.0 throughout. However, they rank below dispersion on each grid and measure only the amount of retrieved support, not neighbor proximity or outcome agreement.

Table~\ref{tab:diagnostics} evaluates all seven diagnostics on the 183,422 ELES records for which every diagnostic is defined, representing 77.34\% of covered records. The restriction arises because weighted \acs{cct} dispersion and neighborhood compactness require at least two retrieved neighbors. The leading diagnostics therefore do not provide a portable confidence score.
\end{sloppypar}

One reason the diagnostics do not transfer is that retrieval's founding assumption, that nearby feature vectors have similar outcomes, is more strongly supported in one archive than the other. Within a fixed location and critical-generator slice, feature distance tracks absolute \acs{cct} difference at Spearman $\rho=0.837$ on BUS39 and $\rho=0.141$ on ELES. Restricting ELES to exact topology matches increases the correlation to 0.380, but only in the small subset with sufficient neighbors. Supplementary Note~S3 reports this diagnostic and verifies that the tested index returns exact neighbors.

\section{Discussion}\label{sec:discussion}

The retrieval method was designed to satisfy three requirements: estimates retain their supporting cases (R1), retrieval abstains outside topology-compatible support (R2), and support statistics are evaluated rather than asserted (R3). The evaluation identifies R2 as the principal operational constraint because compatible support becomes unavailable before conditional accuracy deteriorates substantially.

The consequences of R2 differ between the two archives consistently with their partitioning under the compatibility key and their sampling density; the retrieval rule itself is identical. The archives also differ in dimensionality, physical system, and contingency population, so this pattern is an association rather than an isolated causal factor. BUS39 is densely sampled and forms a single topology group, enabling retrieval for nearly every record. In ELES, one-quarter of records remain unanswered even retrospectively, and coverage decreases by a further 17.13 percentage points when only states available at query time are eligible; conditional error increases by 10.0\% on the records that remain covered. A substantially larger error increase occurs only under the symmetric stress test. Thus, the principal limitation in this archive is compatible coverage rather than the conditional quality of retrieved estimates. Availability could not be evaluated for BUS39 because its sampled states lack acquisition timestamps. BUS39 should therefore be interpreted as a favorable dense benchmark, not as evidence that one archive can satisfy all requirements simultaneously.

These archive results define an operational requirement for deployment: the reference library must acquire states with sufficient temporal density and compatible topologies for queries to identify recent neighbors.

Satisfying the design requirements does not establish that retrieval is preferable to supervised alternatives. On BUS39, where coverage is effectively complete, a standard extra-trees regressor has lower error than retrieval at every reported quantile. On ELES, gradient boosting is more accurate under every tested symmetric exclusion, including the condition that removes the same near-duplicates from its training data (Supplementary Note~S1). Gradient boosting is not more accurate under the one-sided causal rule, where retrieval leads by 4.7\% of its own error on the 181,868 causally covered records, \qty{0.0530}{\second} against \qty{0.0555}{\second}. That comparison is asymmetric in the supervised model's favor, since it is scored without refitting and therefore retains training states recorded after each query (Methods), so the 4.7\% advantage is conservative. Retrieval also has 7.7\% lower error than gradient boosting on the matched ELES population, but this advantage depends on temporally adjacent states and reverses when they are excluded; retrieval has no corresponding advantage on BUS39. In the screening analysis, gradient boosting evaluated at retrieval's referral load with an out-of-fold cutoff is statistically indistinguishable from retrieval. The operational value of retrieval's supporting cases and explicit abstention was not evaluated and would require separate assessment.

Support diagnostics contain information about error, and several rank estimates more effectively than unsorted selection in each archive. However, their relative performance does not transfer: no tested diagnostic ranks first on both grids. None should therefore be interpreted as a portable confidence measure without validation on the deployment archive.

Every reported coverage figure is conditional on exact equality of line status, a compatibility convention rather than a mechanism for improving accuracy. Incorporating 907 binary line indicators into a Euclidean metric alongside continuous voltages and powers would require an explicit scaling between switching changes and continuous electrical quantities. The hard equality constraint avoids this arbitrary scaling and guarantees that each retrieved case shares the query's switching configuration, thereby supporting auditability. This compatibility remains partial: retrieved neighbors commonly differ in generator commitment (Methods). Across three 300-state ELES samples, disabling the filter reduces mean and upper-quantile error on common support while increasing the largest observed error by 4.5 to 22\%; however, a single maximum does not establish a tail bound (Supplementary Tables~S3, S4). The observed fragmentation is also not intrinsic to the assessed system because every varying line flag in this archive represents a neighboring system. Retaining these flags may nevertheless be justified because neighboring switching affects flows within the assessed system. However, after approximately controlling for continuous feature distance, the analysis does not distinguish pairs included by the key from those it excludes (Supplementary Note~S4). The evidence therefore supports a narrower claim than the general proposition that topology fragments operational archives: under the compatibility definition used here, this archive partitions into many short-lived groups, which increases the referral rate in Table~\ref{tab:screening}. Whether a graded alternative, such as learned weighting of switching status or a graph-neural-network topology representation~\cite{qiao2021gnntopology,liu2023gnntsa,liao2022gnnreview}, can retain the compatibility guarantee while reducing coverage loss remains untested.

The retrieval geometry introduces additional limitations. The founding-assumption correlation increases under topology restriction for only 61 ELES queries, so selection may partly explain the increase. Per-column standardization prevents physical units from dominating distance but does not make every feature equally relevant to \acs{cct}; a learned metric could address this limitation~\cite{linjeon2006}.

Four kinds of limitation bear on how far these results carry: the model comparison, evaluation design, ground truth, and study scope. The supervised comparison is limited by comparator selection and the feature budgets used for the tree ensembles. The reported performance difference varies sufficiently among comparators that a comparison with a single baseline partly reflects baseline choice. In addition, every reported row addresses the conditional task of estimating the consequence of a fault at a specified location, as required by a contingency list, rather than unconditional prediction. For the alternative task that also requires fault-location selection, retrieval achieves \qty{0.1312}{\second} on BUS39 and \qty{0.1594}{\second} on ELES, with higher error than a location-median baseline in both datasets. These values characterize a different task and should not be interpreted as deployment estimates. The \qty{0.01}{\second} label grid (Methods) limits the precision with which the results can be interpreted: retrieval's \qty{0.0043}{\second} ELES advantage is less than half one grid step (95\% CI 0.0029 to 0.0056 with states resampled, 0.0023 to 0.0061 with topology groups). The advantage remains after rounding both predictions to this grid, and the two absolute errors differ by at least one step on 70.2\% of records. The small mean difference therefore results from offsetting record-level differences rather than a consistent record-level advantage. None of the models was tuned within the training folds, and all provide point predictions only.

The evaluation design introduces two further limitations (Methods). Preprocessing uses the complete pre-fault feature population, including unlabeled test-state features, and an ELES query may retrieve states recorded before or after its timestamp. The reported results are therefore retrospective and temporally non-causal and do not measure live-deployment performance. The chronological audit quantifies the availability difference: a library frozen at a cutoff leaves almost no subsequent ELES state with a topology-compatible predecessor, and even continuous ingestion provides less support than the retrospective evaluation.

The simulation horizon limits the ground truth rather than the estimator. Multi-swing instability occurring after the simulation horizon would classify a case as stable at a clearing time that it would not survive, producing optimistically biased \acs{cct} labels. The direction of the effect on the recall ratio in Table~\ref{tab:screening} cannot be determined a priori: relabeling a truly unsafe state increases recall if that state would otherwise have been missed and decreases recall if it would have been referred. The effect on the number of missed unsafe states is unfavorable. Retrieval estimates are weighted means of the same optimistically biased labels, so a state whose recorded clearing time exceeds $\tau$ while its true clearing time is below $\tau$ may also receive an above-threshold estimate and remain unreferred. The direction of the net effect on reported recall therefore remains indeterminate without re-simulation at a longer horizon, which was not performed.

Four scope limits remain. The ELES results are derived from one seven-month period in one system and therefore characterize this archive rather than transmission-system archives generally. The term unsafe refers to the minimum \acs{cct} among the evaluated contingencies, which provide broad but non-exhaustive coverage for ELES. The diagnostics characterize retrieval support rather than a calibrated probability of error. Although the underlying assessment system produces frequency- and small-signal-stability outcomes, this study evaluates only transient stability.

Two directions follow. Measured coverage gaps could guide simulation acquisition by identifying the configurations and contingencies for which the estimator must abstain, filling gaps deliberately rather than waiting for a configuration to recur. And every model compared here produces a point prediction, so evaluating the screening trade-off rather than mean error would need quantile regression~\cite{koenker1978regression} or conformal prediction~\cite{vovk2005algorithmic,angelopoulos2023conformal,mollaali2024conformalized}.

The three requirements yield distinct empirical outcomes. R1 and R2 are implemented by design; the evaluation quantifies their consequences and tests the premise of R3. Under causal availability, retrieval covers 56.34\% of records, and paired \acs{mae} on the covered population is 10.0\% higher than under retrospective availability. In the seven-month ELES archive, R2 is satisfied only within short periods of recurring topology, leaving most topology configurations without usable support. Whether another archive can satisfy it is primarily a question about archive acquisition rather than the retrieval algorithm. Much of the answer can be established before fitting an estimator: compatible coverage and time-ordered availability follow from timestamps, switching status, and recorded outcomes interpreted under the adopted compatibility key. The support retained during retrieval additionally depends on the eligibility rule, and the observed accuracy penalty is specific to the implementation evaluated here. The adopted requirements are more likely to be operationally feasible where topology remains stable for long periods or simulation resources deliberately fill coverage gaps. Neither condition holds in the ELES archive.

\section{Methods}\label{sec:methods}

\begin{figure}[tbp]
\centering
\resizebox{\textwidth}{!}{
\newcommand{\subtxt}{\fontsize{7.2}{8.2}\selectfont}
\begin{tikzpicture}[
    font=\fontsize{8.5}{10}\selectfont,
    node distance=3mm and 3mm,
    box/.style={draw, rounded corners=1pt, align=center, fill=gray!10,
                inner sep=4pt, minimum height=10mm, text width=22mm},
    proc/.style={draw, rounded corners=1pt, align=center, fill=blue!8,
                 inner sep=4pt, minimum height=10mm, text width=22mm},
    store/.style={draw, cylinder, shape border rotate=90, aspect=0.12,
                  align=center, fill=orange!15, inner sep=2pt,
                  text width=20mm, minimum height=15mm},
    outbox/.style={draw, rounded corners=1pt, align=center, fill=green!10,
                   inner sep=4pt, minimum height=10mm, text width=22mm},
    lab/.style={font=\subtxt\itshape, gray, fill=white, inner sep=1.4pt},
    arr/.style={-{Latex[length=2mm]}, semithick},
    qarr/.style={arr, dashed, blue!50!black},
]

\node[box] (query) {Query pre-fault state\\(operational snapshot)};
\node[proc, right=of query] (phi) {Same mapping\\$\phi(\cdot)$};
\node[proc, right=of phi] (topo) {Exact topology\\filter};
\node[proc, right=of topo] (knn) {$K$-nearest-neighbor\\similarity search\\(Euclidean)};
\node[proc, right=of knn] (agg) {Record retrieval $+$\\distance-weighted\\aggregation};
\node[outbox, right=of agg] (dash) {CCT estimates,\\outcome distributions,\\neighborhood-support\\diagnostics};

\draw[qarr] (query) -- (phi);
\draw[qarr] (phi) -- (topo);
\draw[qarr] (topo) -- (knn);
\draw[qarr] (knn) -- (agg);
\draw[qarr] (agg) -- (dash);

\node[store, above=10mm of knn] (index) {Feature-vector\\index};
\node[store, text width=22mm, above=10mm of agg] (records)
    {Contingency-record\\archive\\{\subtxt states and outcomes}};

\draw[arr] (index.south) -- (knn.north);
\draw[arr] (records.south) -- (agg.north)
    node[lab, midway, right=1mm] {associated contingency records};

\node[proc, above=10mm of index] (prep) {Feature mapping $\phi$\\[1pt]
    {\subtxt sine--cosine angle encoding,}\\
    {\subtxt per-column standardization}};
\node[proc, left=of prep] (sim) {Offline time-domain\\simulation};
\node[box, left=of sim] (inputs) {Grid model and\\sampled pre-fault states};

\draw[arr] (inputs) -- (sim);
\draw[arr] (sim) -- (prep);
\draw[arr, dashed, gray] ([xshift=-10mm]prep.south) -- ++(0,-5mm)
    -| (phi.north);

\draw[arr] (prep.south) -- (index.north)
    node[lab, pos=0.18, left=1mm] {feature vectors $z_i$};
\draw[arr] (prep.east) -| (records.north)
    node[lab, pos=0.25, above] {records};

\node[lab, anchor=south west] at ([yshift=1mm]inputs.north west)
    {Offline: database construction};
\node[lab, anchor=north west] at ([yshift=-1mm]query.south west)
    {Online: retrieval-based inference (no new simulation, no calibrated intervals)};

\end{tikzpicture}
}
\caption{Retrieval-method architecture. Offline simulations populate a feature index and linked contingency archive. Online, the fitted mapping $\phi$ transforms a query, exact-topology filtering restricts candidates, and distance-weighted retrieval produces \acs{cct} estimates, supporting outcomes, and diagnostics. The query runs no new simulation and provides no calibrated prediction interval.}
\label{fig:schema}
\end{figure}
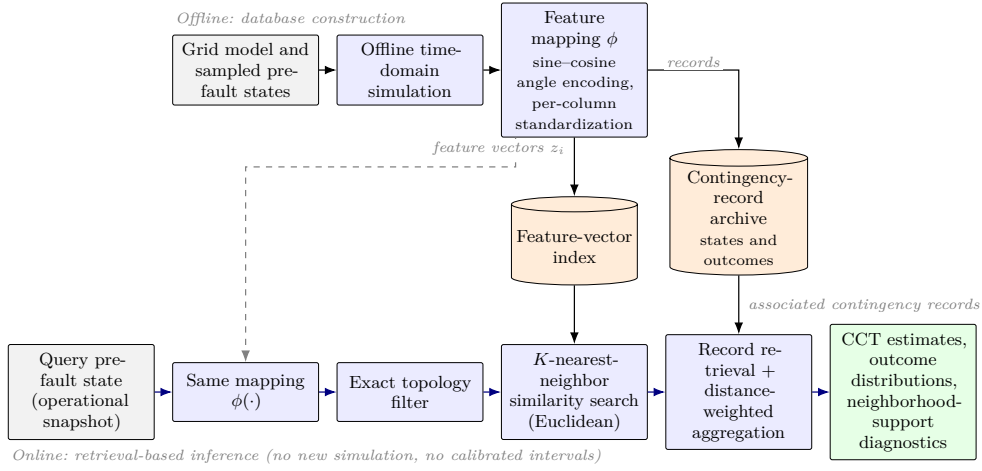

Figure~\ref{fig:schema} separates offline database construction from online retrieval and aggregation.

\subsection{Problem setting}\label{sec:problem-setting}

Each pre-fault state is represented by load-flow variables and has simulated \acs{cct} and critical-generator outcomes for multiple contingencies. The method retrieves states with similar pre-fault conditions, aggregates their known outcomes, and abstains when no compatible reference state exists.

\subsection{Datasets}\label{sec:datasets}

BUS39 uses the standard IEEE New England 39-bus network without topology modifications, with 10 synchronous-machine models, 39 buses, 34 transmission lines, 19 loads, and 12 transformers. Generator G\_01 represents the connection to the external system and acts as the slack generator; the other nine machines represent generators within the modeled grid. Two authors generated the operating states and transient-stability simulations for this study rather than reusing a published dataset. Balanced \acl{rms} (\acs{rms}) simulations were performed in DIgSILENT PowerFactory 2024 SP4S with the full network topology. Operating states were produced from a base case by sampling an overall loading factor from a normal distribution with mean 1 and standard deviation 0.3. The active- and reactive-power setpoints of every generator and load were independently multiplied by factors sampled from a normal distribution with mean 1 and standard deviation 0.1, and then by the overall loading factor. G\_01 absorbed the remaining active-power imbalance during load-flow solution. The study assumes that the grid is stable before a fault occurs. States found to be unstable before fault application therefore failed this eligibility condition and were excluded from further analysis; their exact number is unavailable. The resulting 21,783 retained pre-fault states were each simulated for up to 59 short-circuit contingencies, yielding 1,044,797 state-contingency records.

All simulated faults are solid three-phase short circuits. BUS39 includes bus and line faults, whereas ELES includes line faults only. Each line-fault position was sampled uniformly along the affected line and assigned to its nearest terminal for reporting, reflecting how its location would typically be reported operationally. Bus faults are self-clearing; line faults are cleared by disconnecting both ends of the affected line. A trajectory is classified as unstable when the separation between the rotor-angle changes of any two generators exceeds \ang{180}, where each change is measured relative to that generator's steady-state rotor angle. At the point of detected instability, the critical generator is the generator whose rotor angle lies furthest from the center of inertia,
\begin{equation}
\delta_{\mathrm{COI}}=\frac{\sum_{i=1}^{N}H_i S_{n,i}\delta_i}{\sum_{i=1}^{N}H_i S_{n,i}},
\end{equation}
where $H_i$, $S_{n,i}$, and $\delta_i$ are the inertia constant, nominal apparent power, and rotor angle of generator $i$, respectively. The critical-generator label is taken from the unstable simulation at the clearing time immediately above the \acs{cct}. The simulation stops when instability is detected; otherwise, it continues for up to 4\,s after fault inception. Because the clearing-time search extends to 2.00\,s, the post-clearing observation window narrows as clearing time grows, and multi-swing separation developing later than this horizon would not be detected. Any affected case would be labeled stable at a clearing time that it does not survive, so the \acs{cct} labels would be optimistically rather than conservatively biased. This bias would also affect the recall figures in Table~\ref{tab:screening}. We did not quantify it because doing so requires re-simulating a boundary-stratified sample with a longer horizon. The bisection search likewise assumes that stability is monotonic in fault duration, which we did not test. \acs{cct} is determined by bisection between 0.06 and 2.00\,s until the interval width reaches 0.01\,s, with the longest stable duration retained. Thus, every recorded \acs{cct} lies on a \qty{0.01}{\second} grid, and no label distinguishes clearing times closer than this resolution. PowerFactory's default variable-step trapezoidal solver was used with a base step of 0.01\,s. A result below \qty{0.06}{\second} means that the case is unstable even at the minimum tested clearing time and is operationally classified as absolutely unstable. The raw output encodes these results as 0\,s, and the parser maps them to \qty{0.06}{\second}. These 239 left-censored records, representing 0.02\% of BUS39, have understated regression error. ELES contains no such records. A result above \qty{2.00}{\second} means that the case remains stable at the maximum tested clearing time and is operationally classified as absolutely stable; its true \acs{cct} is right-censored by the search limit. No record in either dataset reaches that limit: the largest recorded \acs{cct} is \qty{1.97}{\second} on BUS39 and \qty{1.95}{\second} on ELES. Thus, no result is right-censored in practice, and the point-error calculations do not treat a bound as an exact outcome. The bound encoding does not affect classification at thresholds from 100 to 200\,ms.

ELES applies the same rotor-angle stability assessment and iterative \acs{cct} procedure to a combined dynamic model of the Slovenian power system and its neighboring systems~\cite{skrlec2025dynamic}. The Faculty of Electrical Engineering and ELES developed the model to represent the actual interconnected system as closely as the available data allowed and jointly validated its simulations. For each assessment, a steady-state observability-area model exported from the ELES energy management system in PSS/E RAW format was merged at the cross-border tie lines with the detailed Slovenian dynamic model and its timestamped topology and setpoints. Overlapping representations of Slovenia and boundary equivalents were removed. Steady-state generators in the observability-area model were replaced by sixth-order synchronous-machine models with governor and excitation models. Generator setpoints outside rated limits were adjusted before dynamic simulation; equivalent loads compensated active-power adjustments and shunt impedances compensated reactive-power adjustments to preserve the original power-flow condition.

After load-flow solution, calculated Slovenian line flows and bus voltages were compared with field measurements. The assessment was aborted if their average deviations exceeded 5\%. A contingency-free time-domain simulation was then used to verify that the initialized model remained stable; detected oscillations also caused the assessment to be aborted. The source dynamic security assessment (DSA) campaign attempted 5,136 operating states between 24 June 2024 and 26 January 2025. Of these, 4,429 (86.23\%) completed successfully and 707 ended with execution errors, including filename or element-name inconsistencies, state-estimator or load-flow non-convergence, and unexpected topology configurations. A further 36 completed states are absent from the delivered tables and are not recovered here, leaving the 4,393 analyzed. Because unexpected topology configurations are among the recorded failure causes, attrition is not independent of switching, and the retained archive may be the one most favorable to exact matching. That campaign and the dynamic model behind it are reported separately~\cite{skrlec2025dynamic}. The present work does not re-report the assessment procedure; it treats the campaign's simulation outputs as the reference library for retrieval-based estimation, which is the subject evaluated here.

The ELES archive used in this study retains 4,393 of the successfully assessed operating states, representing timestamped grid conditions between 27 June 2024 00:15 and 26 January 2025 23:15. ELES provided measurements for every retained timestamp, and these measurements were mapped to the corresponding components of the modeled topology. Each mapped snapshot was required to produce a valid, stable pre-fault simulation state; failure to do so indicated a simulation setup error rather than an operating state within the study scope. States follow a nominal hourly schedule, with gaps in the retained series. The archive maps each analyzed state to a unique acquisition timestamp. Each state has up to 172 simulated line-contingency records, giving 322,823 in total. Under the data agreement, we report only aggregate or anonymized results and name no generator, substation, or line. A superseded 4,402-state batch on the same grid model was used during development but contributes none of the reported results.

These maxima are not reached by every state: BUS39 averages 47.96 records per state and ELES 73.49. Because a shortfall concentrated on severe cases would make every reported error optimistic, we audited its direction against pre-fault loading, a covariate that no simulation outcome can influence. Records per state increase with loading (Spearman $\rho=0.872$ for BUS39 and $\rho=0.363$ for ELES), while observed mean clearing time decreases with it ($\rho=-0.986$ and $\rho=-0.299$; all $p<10^{-90}$). Heavily loaded, severe states therefore carry more records rather than fewer, so the shortfall is concentrated in mild states rather than severe ones. The test is at the state level and does not establish whether the contingencies missing from a given state are the harder ones within it, which remains untested.

Both datasets store \acs{cct} in seconds and a categorical critical-generator label. BUS39 has one topology group because its switching-status flags are constant. Matching ELES's 907 line-status flags gives 1,785 groups of 1 to 38 states, with a median size of one. Including all switching flags, including generator status, instead gives 4,281 groups, with only 4.80\% of states sharing a group with another state. This fragmentation originates outside the assessed system for line status, but not for generator commitment. Of the 165 generator-status flags added by the fuller key, 123 vary across the record. Together, they change state 28,955 times over the 4,392 transitions between consecutive states in time order, and grouping by generator status alone yields 4,153 distinct configurations among 4,393 states. Such a key functions nearly as a unique identifier rather than a compatibility criterion.

Commitment in the neighboring systems represented by the observability-area model follows economic dispatch and is a weak indicator of whether two Slovenian operating states are dynamically comparable. However, most variation does not originate in those neighboring units: matching the 165 flags against ELES's element dictionary identifies 72 Slovenian machines, of which 69 change state during the study period. A smaller Slovenian-specific key collapses this data window to a single group and therefore functions as a no-filter control. This key contains only line and load elements; generators were excluded because domestic commitment changes too frequently to serve as an equality criterion. The absence of generator flags therefore reflects key construction and is not evidence of stable domestic commitment. We matched on line status because the fuller key leaves almost no retrievable support and the narrower key imposes no effective constraint. However, all three keys were explored on the same batch rather than selected through independent validation.

Generator commitment therefore enters through the continuous features rather than the filter. A machine that is not operating has no recorded injection, and the pipeline imputes missing values as zero before standardization; an uncommitted unit is consequently represented by zero active and reactive power. This representation is exact for the power-flow condition but does not fully represent the dynamic state: it does not distinguish a disconnected machine from one synchronized at zero output, although these conditions differ in inertia and voltage support and can therefore affect clearing time. Matched neighbors can consequently agree on line status and injected power while differing in the number of synchronized machines. In this archive, such differences occur in nearly all matched pairs. Of 10,697 state pairs sharing a line-status group, sampled at up to 400 per group from 11,230, 98.80\% differ in generator commitment, with a median difference of 11 of the 123 varying flags.

For pairs describing the same fault location and critical generator, absolute \acs{cct} differences average \qty{0.0205}{\second} when commitment also agrees and \qty{0.0584}{\second} when it does not. These differences increase with the number of machines for which commitment differs (Spearman $\rho=0.281$). For reference, retrieval's ELES error is \qty{0.051}{\second}, corresponding to the five-fold matched-comparison \acs{mae} in Table~\ref{tab:extratrees}. This pairwise association is not a causal estimate because states with different commitment also differ in dispatch and loading, and the 128 commitment-identical pairs are likely to be similar in these respects. Nevertheless, the result identifies a limitation of the line-status key: retrieved neighbors are compatible in switching status but commonly differ in generator commitment.

\subsection{Feature representation}\label{sec:feature-representation}

The feature mapping $\phi$ replaces phase angles with their sine and cosine to avoid the discontinuity at \ang{\pm 180} and standardizes every other numeric column to zero mean and unit variance. The resulting vector has 260 dimensions on BUS39 and 12,524 on ELES. Every switching-status column is excluded from the searched vector, 44 on BUS39 and 1,072 on ELES, leaving 216 and 11,452 dimensions respectively, so the two datasets differ in retrieval geometry as well as in topology fragmentation. Those excluded columns do not all serve the same purpose. Of the 1,072 on ELES, only the 907 line-status columns form the deployed exact-equality key; the 165 generator-status columns are excluded from the distance and unused by the filter, so generator commitment enters neither, and is represented only indirectly through the active and reactive injections of the machines concerned. The scaler uses the complete pre-fault feature population, including unlabeled test-state features, so preprocessing is transductive. A prospective evaluation should fit preprocessing only on its historical reference period.

\subsection{Retrieval}\label{sec:retrieval}

A query is compared only with reference states that share its switching-status configuration. Nearly identical topologies are treated as incompatible. Within this pool, the method retrieves the $K$ nearest states in Euclidean feature distance, with $K=100$ unless stated otherwise. If fewer than $K$ matching states exist, all are returned. If none exist, the system abstains rather than using a different topology.

The guarantee is therefore over switching configurations and not over the continuous features: a query far from every state in its own group is still answered, since no distance threshold is applied. On ELES, $K=100$ never limits retrieval because topology groups contain at most 38 states. After excluding the query, every ELES search uses its complete same-topology pool. On BUS39, $K$ selects a subset from the single group of 21,783 states. The prototype uses Qdrant in embedded (\texttt{:memory:}) mode. At these collection sizes, it may perform an exact scan. The index-fidelity check below confirms agreement with brute-force search for the tested queries but does not validate approximate \acl{hnsw} (\acs{hnsw}) retrieval in production.

\subsection{Retrieval-quality validation}\label{sec:retrieval-quality}

The founding-assumption and index-fidelity analyses test retrieval itself rather than the aggregated \acs{cct} estimate. We sample 300 query states per dataset without replacement using seed 42. For the founding-assumption diagnostic, each query uses the (location, recorded critical generator) pair from its first contingency record in stored dataset order. This stored-order choice was not sensitivity-tested. The unrestricted analysis compares feature distance and absolute \acs{cct} difference against every other state with that same pair, regardless of topology. This diagnostic measures distance over the full feature mapping, including the switching-status columns that retrieval itself excludes: on BUS39 those columns are constant and contribute nothing, whereas on ELES they enter the distance. It retains queries with at least 200 candidates on BUS39 or 50 on ELES. All 300 BUS39 queries and 294 ELES queries meet this requirement. The remaining six ELES queries have too few candidates. We report Spearman correlation, calculated as Pearson correlation of the ranks, and compare mean absolute \acs{cct} difference for the nearest 100 states with a random sample of 100 states from the same candidate population.

The topology-restricted analysis additionally requires an exact topology match. It uses $k=100$ on BUS39 and $k=5$ on ELES. The smaller ELES value leaves enough queries for analysis and is not the largest possible group size, which is 37 after excluding the query. We apply this check to the original 300 queries and use an independent random stream with seed 43. Queries with fewer than $k$ candidates are skipped and counted, leaving 61 ELES queries at $k=5$. That eligibility condition is stricter than the index-fidelity one below: it requires $k$ candidates sharing both the query's topology and its (location, recorded critical generator) slice, whereas index fidelity requires only $k$ same-topology candidates.

For index fidelity, queries pass through the production \texttt{EstimationService} and Qdrant path. We compare the returned neighbor set with exact brute-force top-$k$ search on the same topology group, then calculate mean recall@$k$ for queries with at least $k$ candidates. Unlike the founding-assumption diagnostic above, the brute-force reference here is computed over exactly the vector the index holds, so both paths use the same columns, preprocessing, topology pool and distance. This check draws its own sample of 300 query states rather than reusing the set above. It retains all 300 BUS39 queries at $k=100$ but only 130 of the 300 ELES queries at $k=5$, the remaining 170 having too few same-topology candidates. Recall of 1.0 means that the embedded backend matched brute force for every tested query. It does not establish recall for an approximate \acs{hnsw} index.

\subsection{Aggregation and diagnostics}\label{sec:aggregation}

Retrieved states are grouped by recorded fault location and critical generator. Within a group, an exponential kernel converts query distance to weight, $w_i \propto \exp(-\alpha d_i)$ with $\alpha=1$, and the weights are normalized to sum to one. A single-fold BUS39 sweep over $\alpha \in \{0.5,1,2\}$ and $K \in \{25,50,100,200\}$ places the reported setting within the tested range (Supplementary Table S2). The sweep did not select this setting. \acs{cct} is the weighted mean. At each location, candidate generators are ranked by unnormalized kernel-support mass, $\sum_i \exp(-\alpha d_i)$. This quantity measures retrieved support and is not a calibrated probability of becoming critical.

The system also reports diagnostics of each retrieved group. None is presented as calibrated confidence. Effective sample size,
\begin{equation}
n_\mathrm{eff} = 1 \Big/ \sum_i w_i^2
\end{equation}
indicates whether a few neighbors dominate the weights. Neighborhood compactness,
\begin{equation}
\frac{1}{\binom{n}{2}} \sum_{i<j} \exp(-\alpha \, d_{ij}),
\end{equation}
is the mean kernel value over all unique pairwise distances $d_{ij}$ among the $n$ retrieved neighbors. It describes similarity within the neighborhood and does not include the query. Weighted \acs{cct} dispersion,
\begin{equation}
\mathrm{cct\_weighted\_std} = \sqrt{\textstyle\sum_i w_i \left(\mathrm{CCT}_i - \overline{\mathrm{CCT}}\right)^2}, \quad \overline{\mathrm{CCT}} = \textstyle\sum_i w_i \, \mathrm{CCT}_i,
\end{equation}
measures agreement among retrieved outcomes. Input similarity and outcome agreement need not coincide, so their relationship with error is tested in the Results. At $n=1$, weighted dispersion would be zero but cannot demonstrate agreement, while compactness has no neighbor pair and is undefined. The implementation therefore returns both as undefined below $n=2$.

One further candidate was computed and not reported, and we describe it here so that the reported set is not mistaken for a selection made on performance. The weighted Pearson correlation between neighbor distance and \acs{cct}, $\mathrm{cct\_distance\_correlation}$, would rank support by its absolute value, but it is undefined for groups with fewer than two neighbors or with no variation in distance or \acs{cct}. That affects more than one-third of ELES groups, so including it would have shrunk the population on which all the others are compared. It was excluded on availability, before its \acs{naurc} was examined.

\subsection{Conditional critical-generator evaluation}\label{sec:deoracling}

For each contingency, the recorded fault location selects the corresponding retrieval report. The oracle estimate uses the recorded critical generator, while the selected-generator estimate uses the candidate with the greatest kernel-support mass. The location is known in both rows, and only the generator differs. Because the two rows cover different populations, we report their MAEs with sample sizes and do not treat the marginal difference as a selection cost. We additionally restrict both levels to their common support and compare them pairwise there. The paired record-level \acs{mae} difference uses a state-clustered bootstrap, and a paired Wilcoxon signed-rank test on per-state mean absolute errors gives the corresponding significance, since records within a pre-fault state are not independent. Effect sizes are reported as $r=z/\sqrt{n}$ with $z$ signed by direction, positive when selection is worse than the oracle, and $n$ the number of states with a non-zero paired difference rather than the number of states compared. Both datasets reject equality: BUS39 $r=0.643$ and ELES $r=-0.130$, with the opposite signs reflecting the reversal described in the Results.

Top-1 identification is the fraction of records with a candidate at the recorded location for which the highest-support generator matches the recorded generator. Records where the generator is absent from the candidate set are reported separately as coverage gaps.

The supervised models use the same globally fitted feature scaler, one-hot encoded contingency location, and switching flags as numeric features rather than a hard filter. They exclude the recorded generator, fault terminal, fault type, \acs{cct}, and state identifier. Three supervised models are compared: extra trees and a random forest, each with 300 trees, \texttt{min\_samples\_leaf=2}, and \texttt{random\_state=42}, and a histogram gradient-boosting regressor with absolute-error loss, 300 iterations, and learning rate 0.05. Five-fold \texttt{GroupKFold} cross-validation groups records by pre-fault state. Within each training fold, a variance threshold removes constant columns. The per-split feature budget of the tree ensembles is all features on BUS39 and $\sqrt{p}$ on ELES, the latter chosen for tractability on that dataset's 12,524-dimensional representation; gradient boosting has no such parameter. A location-median baseline is fitted within each training fold.

Results were produced with Python 3.12.13, scikit-learn 1.9.0, qdrant-client 1.17.1, NumPy 2.4.6, pandas 3.0.3 and SciPy 1.17.1, and PowerFactory 2024 SP4S for the simulations. The tree-ensemble rows are sensitive to the scikit-learn build: an earlier version of this work fitted them through the \texttt{scikit-\allowbreak{}learn-\allowbreak{}intelex} extension, whose substituted implementations are not numerically equivalent to upstream scikit-learn. The unmodified library reports 14 to 48\% higher error on those $\sqrt{p}$ rows. The rows above use unmodified scikit-learn throughout.

The temporally blocked variant additionally removes from each training fold every record whose state lies within three hours of any test-fold state, so that the supervised model is denied the same near-duplicates the retrieval control withholds. Only the three-hour window is reported: grouped folds assign states to folds without regard to time, so test states are spread across the whole record and a wider exclusion leaves almost no training data at all. Comparing at wider windows would require time-contiguous folds, which we did not run.

For the matched comparison in Table~\ref{tab:extratrees}, retrieval uses the same five folds and excludes every state in the test fold from its reference library. It then selects the candidate generator with the greatest kernel-support mass at the recorded fault location. The supervised models and the location-median baseline are restricted to the same retrieval-covered records: 1,044,792 BUS39 records and 230,511 ELES records. Full-population supervised and location-median errors are calculated separately. This five-fold ELES retrieval run is used only for the supervised comparison; the other ELES analyses remove one query state at a time from the reference population.

\subsection{Operational screening}\label{sec:screening}

A deployment screen refers a state when any evaluated contingency may fall below a clearing-time threshold $\tau$. Following prior work~\cite{pertl2017decision}, the primary threshold is $\tau=\qty{200}{\ms}$, with 100 and 150\,ms as sensitivity values. A state is observed unsafe if its lowest recorded \acs{cct} falls below $\tau$.

For each contingency, the screen uses the minimum weighted \acs{cct} across all candidate critical generators available at that location. It cannot use the recorded critical generator because that label is an outcome of simulation. This conservative estimate is specific to the screening analysis. The oracle results use the recorded generator, while the diagnostic analysis uses the highest-support generator. The coverage figures require only that some candidate exists at the location.

A state is referred to time-domain simulation if its minimum available estimate is below $\tau$ or if any recorded contingency has no estimate. The latter rule treats incomplete retrieval as abstention. Recall is the share of observed-unsafe states referred, false-alarm rate is the share of other states referred, and state-referral rate is the share of all states referred.

\subsection{Temporal-exclusion control}\label{sec:temporal-exclusion}

Two controls restrict the states eligible for retrieval. The causal control includes only states recorded at or before the query timestamp minus an ingestion lag and always excludes the query itself; this represents the rule available in deployment. We report lags of zero and one hour. The symmetric control excludes every state whose timestamp lies within a specified window on either side of the query. This control additionally removes recent history that would be available in deployment and therefore provides a conservative assessment of dependence on temporal proximity rather than a deployment model. The window is evaluated at 3, 6, 12, 24, and 48 hours, with separate output generated for each run. All other settings, including the topology filter, feature mapping, $\alpha$, and highest-support generator selection, remain unchanged. The retrieval path verifies that no excluded state appears among the returned neighbors. We compare the two runs on records that receive an estimate under both conditions, yielding a paired accuracy comparison; coverage loss is reported separately. Confidence intervals use 1,000 state-level bootstrap resamples, and significance is evaluated using a paired Wilcoxon signed-rank test on per-state mean absolute errors, consistent with the analyses above. The supervised baselines are evaluated on the same surviving records rather than on their full populations, and one model is additionally refitted under the matched temporal-exclusion condition described above. This refit applies only to the symmetric control. Under the causal control, the supervised model is evaluated without refitting and therefore retains training states recorded after each query, whereas retrieval cannot use them. The causal comparison is consequently conservative for retrieval rather than fully matched.

\subsection{Chronological availability audit}\label{sec:chronology-audit}

This audit uses state timestamps and topology keys only. It runs no retrieval and reproduces no estimate. For each ELES state we test whether another state sharing its topology key has a strictly earlier timestamp, under two availability scenarios. The expanding scenario allows any strictly earlier state and models a reference library that ingests outcomes continuously without delay. The frozen scenario restricts candidates to states before a cutoff, models a static library, and evaluates only the states after that cutoff. The cutoffs are the nine deciles of elapsed calendar time and were fixed before any support value was computed, so they are not selected on the results. The same timestamps give each topology group's calendar span, the interval between its first and last state, and its burst count, the number of clusters left after splitting the group's sorted timestamps wherever consecutive states lie more than 24 hours apart. That threshold is a reporting choice rather than a fitted one; states are sampled hourly, so any value above an hour separates distinct occasions rather than sampling jitter, and results for 6 to 72 hours accompany the released data. Groups holding one state have no span and are counted separately.

We report the deployed line-status key. The all-flag key gives 2.55\% support under continuous ingestion, and the coarse Slovenian key, which collapses this data window to a single group, gives 99.98\% and therefore functions as a no-filter control. A deployment cannot use a state immediately after it is recorded because its contingencies must first be simulated; we therefore evaluated whether this delay changes availability. No change occurs for structural reasons: states are sampled hourly and the cutoff is inclusive, so any lag up to one hour removes no candidate retained in the zero-lag scenario. Only a delay exceeding the sampling interval would alter these figures. The audit assumes immediate insertion and sufficient parallel capacity to prevent contingency queues, representing a best-case assumption rather than an observed processing cadence. Because the test is conducted at the state level, it does not establish the availability of any specific contingency outcome. The audit therefore provides design evidence for a prospective study rather than a substitute for one.

\subsection{Statistical evaluation}\label{sec:stat-evaluation}

Both datasets are evaluated leave-one-state-out: a query retrieves from every state but itself, so no record is scored against its own operating state. The one exception is Table~\ref{tab:extratrees}, where retrieval and the supervised models share five \texttt{GroupKFold} splits by pre-fault state, because a regressor would otherwise need one refit per state. Evaluating retrieval under both protocols quantifies the effect of this exception: withholding a whole fold instead of one state increases retrieval error by 1.0\% on BUS39 and 2.3\% on ELES. The shared-fold protocol therefore does not materially favor or disadvantage retrieval. ELES removes the query state but allows all other states, regardless of timestamp. A reference state may therefore precede or follow the query. This evaluation is retrospective and temporally non-causal rather than chronological or live. Transductive preprocessing is a separate limitation: reference selection concerns which states may be retrieved, while scaler fitting concerns which feature distributions inform preprocessing. No estimate uses its own contingency outcome.

We summarize diagnostic value by \acs{naurc} at 50, 70, 80, 90, 95, and 100\% coverage. Records are ordered from most to least trusted according to each diagnostic, and \acs{mae} is recalculated as the least-trusted records are withheld. If a coverage boundary cuts through records with the same diagnostic value, all better-valued groups are included and the boundary group contributes fractionally to reach the target count. This gives the expected \acs{mae} under uniform selection within that tied group and makes the result independent of row order. Trapezoidal integration gives the area under the six points. We normalize this area by full-coverage \acs{mae} multiplied by the 0.5 to 1.0 coverage span. A value of 1.0 represents the unsorted baseline, values below 1.0 are favorable, and values above 1.0 indicate reversed ordering.

All diagnostics in Table~\ref{tab:diagnostics} are evaluated on the records where every one of them is defined. This retains 1,044,761 of 1,044,793 covered BUS39 records, representing all 21,783 states, and 183,422 of 237,168 covered ELES records, representing all 3,344 covered states. Using common support avoids comparing diagnostics on different record populations when some are undefined for small groups.

Bootstrap resampling occurs at the pre-fault-state level, keeping all contingencies from a state together. Diagnostic intervals use 1,000 resamples, conditional-generator intervals use 1,000, the paired common-support comparison uses 1,000, and operational-screening intervals use 2,000. We report the 2.5th and 97.5th percentiles with random seed 42.

The topology ablation uses three 300-state ELES samples (seeds 42, 123, and 7) and pairs records covered with and without exact filtering. Per-state mean differences are compared by paired Wilcoxon signed-rank tests and 1,000-resample state-bootstrap CIs. Supplementary Table S3 reports effect sizes and maxima, and Supplementary Table S4 the corresponding error quantiles.

\subsection{Use of AI-assisted tools}\label{sec:ai-disclosure}
Claude Code (Anthropic) and Codex (OpenAI) assisted with language editing, code review, consistency checks, methodological critique, and drafting and testing analysis code. The authors reviewed and verified all assisted analyses, results, references, and manuscript changes, made all scientific decisions, and retain full responsibility for the work.

\subsection{Code availability}\label{sec:code-availability}
\begin{sloppypar}
The implementation, the data-preparation pipeline and the analysis scripts that produce every table and figure reported here are available at \url{https://github.com/sensorlab/AI-ASSIST}, at release \texttt{v1.1.0} (commit \texttt{1cb6c90e\allowbreak 69178eea\allowbreak 5a928216\allowbreak 37b8810b\allowbreak f4bd2544}). The analysis scripts are in \texttt{scripts/paper/} and the summary tables and figure sources they write are committed under \texttt{results/}.
\end{sloppypar}

\backmatter

\bookmarksetupnext{level=1}
\bmhead{Supplementary information}
Supplementary Note~S1 and Supplementary Fig.~S1 report the symmetric temporal-exclusion analysis, Supplementary Note~S2 reports the oracle and common-support critical-generator comparison, and Supplementary Note~S3 reports the founding-assumption and index-fidelity checks. Supplementary Table S1 reports absolute \acs{cct} error quantiles for the two conditional critical-generator evaluation levels on both datasets. Supplementary Table S2 reports BUS39 sensitivity to kernel bandwidth $\alpha$ and neighbor count $K$. Supplementary Table S3 reports the topology-filter ablation across all three tested random seeds, and Supplementary Table S4 reports the corresponding error quantiles. Supplementary Note~S4 and Supplementary Table S5 evaluate the contribution of exact line-status equality in the ELES archive.

\bookmarksetupnext{level=1}
\bmhead{Acknowledgements}
This work was co-funded by the Slovenian Research and Innovation Agency (ARIS) under Grants \mbox{No.\,P2-0016} and \mbox{No.\,L2-50053} (AI-ASSIST), and by the EU Horizon Europe Grants \mbox{No.\,101172887} (EnerTEF) and \mbox{No.\,101173007} (InterSCADA).

\bookmarksetupnext{level=1}
\bmhead{Data Availability}
The BUS39 dataset is publicly available at \url{https://github.com/sensorlab/AI-ASSIST}. ELES provided the operating-state and grid-model data, which are not publicly available because ELES classifies them as sensitive critical-infrastructure information. Urban Rudež serves as the project lead, domain expert, and supervisor coordinating collaboration between the participating organizations; requests for an access agreement or authorization should be directed to him at urban.rudez@fe.uni-lj.si. Access remains subject to ELES approval and applicable data-use conditions. The aggregate results reported in the manuscript contain no named ELES grid entities.

\bookmarksetupnext{level=1}
\bmhead{Author Contributions}
\textbf{Gregor Cerar:} Conceptualization, Methodology, Software, Validation, Formal analysis, Investigation, Data curation, Visualization, Writing: original draft, Writing: review and editing. \textbf{Denis Sodin:} Formal analysis, Validation, Writing: review and editing. \textbf{Matjaž Škrlec:} Resources, Data curation, Investigation, Validation, Writing: review and editing. \textbf{Urban Rudež:} Funding acquisition, Project administration, Supervision, Investigation, Writing: review and editing. \textbf{Mihael Mohorčič:} Funding acquisition, Resources, Project administration, Supervision, Writing: review and editing. \textbf{Miha Smolnikar:} Funding acquisition, Resources, Project administration, Supervision, Writing: review and editing.

\bookmarksetupnext{level=1}
\bmhead{Competing Interests}
The authors declare no competing financial interests or personal relationships that could have appeared to influence the work reported in this paper.

\bibliography{biblio}

\clearpage
\section{Supplementary Information}\label{sec:supplementary-information}
\suppressfloats[t]
\setcounter{figure}{0}
\setcounter{table}{0}
\renewcommand{\theHfigure}{S\arabic{figure}}
\renewcommand{\theHtable}{S\arabic{table}}

\renewcommand{\thetable}{S\arabic{table}}
\renewcommand{\thefigure}{S\arabic{figure}}

\paragraph{Supplementary Note S1: symmetric temporal-exclusion sweep.}
Mean absolute error is abbreviated MAE throughout.

This note reports the symmetric exclusion analysis summarized in the main text. The analysis excludes states on both sides of the query and therefore constitutes a stress test rather than a deployment model: it removes recent history that would legitimately be available in deployment and provides a conservative assessment of dependence on temporally adjacent states. Excluding only the query state leaves states recorded immediately before and after it available for retrieval. Because ELES topology groups are temporally concentrated, a query may therefore be answered using an operating state recorded one hour earlier. We repeated the ELES evaluation after excluding every state within windows ranging from 3 to 48 hours around the query (Methods).

\begin{figure}[tbp]
\centering
\definecolor{okabeBlue}{HTML}{0072B2}
\definecolor{okabeVermillion}{HTML}{D55E00}
\definecolor{okabeGreen}{HTML}{009E73}
\begin{tikzpicture}
\begin{groupplot}[
  group style={group size=2 by 1, horizontal sep=1.35cm},
  width=0.47\textwidth, height=5.2cm,
  xmode=log, log basis x=2, xmin=2.2, xmax=66,
  xtick={3,6,12,24,48}, xticklabels={3,6,12,24,48},
  xlabel={Exclusion window (hours, $\pm$)},
  tick label style={font=\scriptsize}, label style={font=\scriptsize},
  title style={font=\scriptsize\bfseries},
  grid=major, grid style={gray!22, line width=0.3pt},
  every axis plot/.append style={line width=0.9pt, mark size=1.7pt},
  legend style={draw=none, fill=none, font=\scriptsize, legend columns=3,
                /tikz/every even column/.append style={column sep=10pt}},
]
\nextgroupplot[title={Support}, ylabel={Records with an estimate (\%)}, ymin=0, ymax=80]
\addplot[gray!65, dashed, line width=0.7pt, mark=none, forget plot] coordinates {(2.2,73.47) (66,73.47)};
\addplot[okabeGreen, dash dot, line width=0.8pt, mark=none, forget plot] coordinates {(2.2,56.34) (66,56.34)};
\addplot[color=okabeBlue, mark=*, mark options={fill=okabeBlue}, forget plot] coordinates {(3,41.14) (6,23.54) (12,12.92) (24,8.37) (48,2.43)};
\nextgroupplot[title={Error on supported records}, ylabel={MAE (s)}, ymin=0.04, ymax=0.115, legend style={at={(-0.14,-0.30)}, anchor=north}, scaled y ticks=false, ytick={0.04,0.06,0.08,0.10}, y tick label style={/pgf/number format/fixed, /pgf/number format/precision=2}]
\addplot[gray!65, dashed, line width=0.7pt, mark=none] coordinates {(3,0.05064) (6,0.05321) (12,0.05358) (24,0.05127) (48,0.05613)};
\addplot[okabeGreen, dash dot, line width=0.8pt, mark=none] coordinates {(2.2,0.05301) (66,0.05301)};
\addplot[color=okabeBlue, mark=*, mark options={fill=okabeBlue}, error bars/.cd, y dir=both, y explicit] table[x=x,y=y,y error plus=ep,y error minus=em] {
    x y ep em
    3 0.08948 0.00261 0.00243
    6 0.09457 0.00314 0.00328
    12 0.09274 0.00385 0.00398
    24 0.09155 0.00449 0.00482
    48 0.10514 0.01005 0.00967
  };
\addplot[color=okabeVermillion, mark=diamond*, mark options={fill=okabeVermillion}] coordinates {(3,0.05509) (6,0.05385) (12,0.05457) (24,0.05569) (48,0.06191)};
\legend{{Query excluded only},{Causal, deployment rule},{Retrieval},{Gradient boosting}}
\end{groupplot}
\end{tikzpicture}
\caption{Effect of withholding reference states recorded close in time to the query, ELES. The left panel shows the proportion of 322,823 records that receive an estimate, and the right panel shows mean absolute error on those records, with gradient boosting evaluated on the same populations. Gray denotes the evaluation that excludes only the query; green denotes the causal rule that permits only earlier states. Error bars are 95\% bootstrap intervals for the paired difference from each window's baseline. The analysis retains 1,959 states at 3 hours and 109 at 48 hours.}
\label{fig:temporal-decay}
\end{figure}
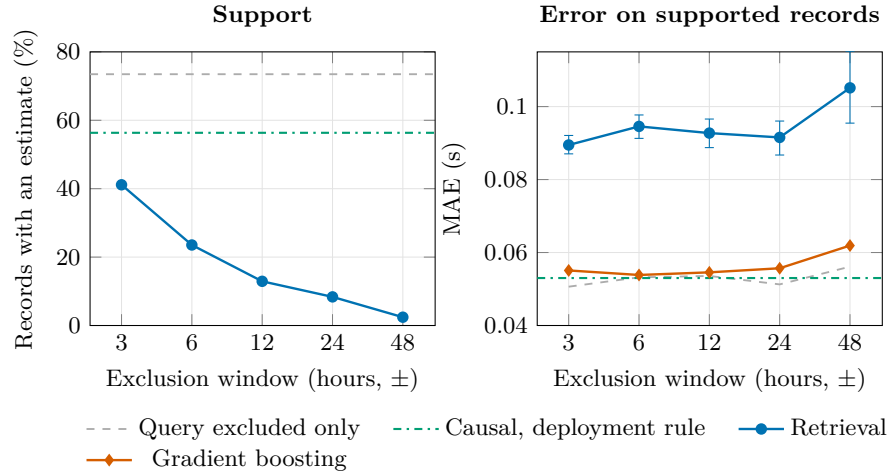

Support and accuracy respond differently to temporal exclusion (Fig.~\ref{fig:temporal-decay}). Coverage decreases from 73.47\% of records when only the query is removed to 41.14\% at 3 hours, 23.54\% at 6 hours, 12.92\% at 12 hours, 8.37\% at 24 hours, and 2.43\% at 48 hours. Error does not change in the same graded manner. Relative to the same records evaluated without temporal exclusion, MAE increases by 76.7\% at the 3-hour window, reaching \qty{0.0895}{\second}, and stays near this level at every wider window, though each window is evaluated on its own surviving records against its own baseline, so these are separate paired comparisons rather than a series and the support shrinks from 1,959 states at 3 hours to 109 at 48: 77.7\% at 6 hours, 73.1\% at 12 hours, 78.6\% at 24 hours, and 87.3\% at 48 hours. Thus, almost the entire error increase occurs when states within three hours on either side of the query are withheld.

The records retained at each window form a selected subset whose topologies recur beyond that window. Under the original protocol, retrieval MAE on these subsets ranges from \qty{0.0506}{\second} to \qty{0.0561}{\second}, compared with \qty{0.0497}{\second} on the complete covered population. The retained subsets therefore have slightly higher baseline error. Each window is compared with its own baseline, so this difference does not affect the relative increases above. The analysis provides a conservative assessment of temporal dependence but does not identify its precise onset because the narrowest tested window already exceeds the hourly sampling interval.

The supervised regressors in Table 3 of the main text, evaluated at the same recorded locations, do not use retrieval; the exclusion analysis therefore leaves their training data unchanged. Gradient-boosting MAE ranges from \qty{0.0539}{\second} to \qty{0.0619}{\second} across the retained populations and is lower than retrieval MAE at every window. At the 24-hour window, retrieval error is 1.64 times that of gradient boosting, compared with 0.92 times when temporal neighbors are available.

This comparison is asymmetric. Grouped folds withhold complete states, but a state recorded one hour earlier is treated as a different state and may remain in the training set. Consequently, 99.4 to 100.0\% of test states per fold have a training state within three hours, and the supervised training sets retain the temporally adjacent states excluded from retrieval.

We therefore refitted gradient boosting using training folds that also exclude states within three hours of every test state (Methods), removing 73 to 76\% of the training records. On the 132,817 records retained after retrieval's three-hour exclusion, the temporally blocked model achieves \qty{0.0598}{\second}, compared with \qty{0.0551}{\second} for the unblocked model on the same subset, an increase of 8.6\%. Retrieval error increases by 76.7\% on these records. Both methods therefore lose accuracy when temporally adjacent states are withheld, but the relative increase for retrieval is approximately nine times larger. This difference indicates that retrieval's dependence on these states is a property of the method rather than solely of the evaluation design. Under the matched temporal-exclusion condition, the blocked supervised model remains more accurate than retrieval, with MAEs of \qty{0.0598}{\second} and \qty{0.0895}{\second}, respectively.

Retrieval accuracy in this archive therefore depends primarily on states recorded within a few hours of the query rather than on the broader reference population. States beyond this interval contribute to coverage but provide limited additional accuracy.

\paragraph{Supplementary Note S2: oracle and common-support critical-generator comparison.}
Among records with at least one candidate at the recorded fault location, the candidate set included the recorded critical generator in 99.96\% of BUS39 cases and 86.62\% of ELES cases. When it was present, the highest-support candidate matched it in 95.5\% and 81.8\% of cases, respectively. Combining retrieval and selection gives success rates of 95.5\% for BUS39 and 70.8\% for ELES among covered records. Across all records, including those without a location-specific estimate, the rates are 95.5\% and 52.0\%. Critical-generator retrieval and support-based selection were therefore less successful on ELES. These results do not isolate the predictive information in fault location because candidate availability and feature-space retrieval also affect success. Table 2 of the main text reports critical clearing time (CCT) errors on each row's covered population, and Supplementary Table S1 gives the corresponding absolute-error quantiles.

Because the rows cover different populations, we also compare them on their common support. On BUS39, this comparison agrees with the marginal result: selection MAE is 4.34\% higher than oracle MAE, and selection has higher error in 78.4\% of states. On ELES, the direction reverses: selection MAE is 2.64\% \emph{lower} than oracle MAE on records covered by both rows, so the oracle is not an error lower bound in this dataset. The marginal comparison measures a different quantity. The two ELES rows differ by 31,735 records whose recorded critical generator does not appear among the retrieved candidates. Excluding these records only from the oracle row produces a systematically easier oracle population. This population difference explains the marginal difference but not the paired result because these records are absent from both conditions on common support.

\paragraph{Supplementary Note S3: founding-assumption and index-fidelity checks.}
Retrieval assumes that nearby feature vectors have similar outcomes. In an outcome-conditioned diagnostic, feature distance correlates with absolute CCT difference at Spearman $\rho=0.837$ on BUS39 and $\rho=0.141$ on ELES within a fixed (location, recorded critical generator) slice for each sampled query. The unrestricted analysis retains 300 BUS39 queries and 294 ELES queries. Restricting ELES candidates to exact topology raises $\rho$ to 0.380 at $k=5$, but only 61 of the original 300 queries have five such neighbors. This higher correlation therefore describes a selected, better-supported subset. Compared with random states from the same slice, nearest-neighbor averaging reduces mean $|\Delta\mathrm{CCT}|$ by 84.0\% on BUS39, 25.2\% on unrestricted ELES, and 18.0\% on the topology-restricted ELES subset. This outcome-conditioned, single-contingency diagnostic describes within-slice geometry, not oracle-free retrieval quality or robustness to contingency choice. The embedded vector index returns the same top-$K$ neighbors as brute-force search for every tested query. This verifies the tested retrieval path, but not approximate-index recall in a production hierarchical navigable small world (HNSW) deployment (Methods).

\paragraph{Critical-generator evaluation.}
For each query at a recorded fault location, retrieval produces a separate CCT
estimate for every critical-generator group represented among the supporting
cases. The recorded critical generator is the generator whose rotor angle lies
farthest from the inertia-weighted center of angles at the point where the
time-domain simulation detects instability, as defined in the Methods. Because this is a
simulation outcome, using its group to choose the CCT estimate is an oracle
evaluation. The non-oracle alternative instead uses the candidate generator
with the greatest unnormalized kernel-support mass,
$\sum_i \exp(-\alpha d_i)$, which favors groups with more and/or closer
supporting cases. This mass ranks the retrieved candidates; it is not a
calibrated probability that a generator will become critical. In both
evaluations, the recorded fault location and target CCT are unchanged.

\begin{tableorg}[htbp]
\centering
\caption{Absolute CCT error quantiles (seconds) for the two conditional critical-generator evaluation levels defined above. The recorded fault location is fixed in both rows. Each row uses its own covered population, matching Table 2 in the main text, because a candidate generator can be available even when the recorded critical generator is absent from the retrieved cases. The row differences are therefore not a paired estimate of generator-selection cost. $q_{100}$ is the maximum observed absolute error.}
\label{tab:deoracling-quantiles}
\resizebox{\textwidth}{!}{%
\begin{tabular}{llrrrrrrrr}
\toprule
& Level & $q_{0}$ & $q_{25}$ & $q_{50}$ & $q_{75}$ & $q_{90}$ & $q_{95}$ & $q_{99}$ & $q_{100}$ \\
\midrule
\multirow{2}{*}{BUS39} & Recorded critical generator (oracle) & 0.000 & 0.006 & 0.013 & 0.023 & 0.037 & 0.050 & 0.108 & 0.791 \\
& Highest-support candidate generator & 0.000 & 0.006 & 0.013 & 0.024 & 0.039 & 0.054 & 0.119 & 0.791 \\
\midrule
\multirow{2}{*}{ELES} & Recorded critical generator (oracle) & 0.000 & 0.000 & 0.010 & 0.040 & 0.117 & 0.199 & 0.490 & 1.400 \\
& Highest-support candidate generator & 0.000 & 0.000 & 0.010 & 0.050 & 0.130 & 0.220 & 0.580 & 1.620 \\
\bottomrule
\end{tabular}
}
\end{tableorg}

\begin{table}[htbp]
\centering
\caption{BUS39 sensitivity of mean absolute error (s) to retrieval-kernel bandwidth $\alpha$ and neighbor count $K$ on one held-out fold of a five-fold split by pre-fault state. The sweep predates the library upgrade that reassigned folds, so its held-out states are not identical to any fold used in the main text; it is a non-selective sensitivity check rather than a tuning result, so the comparison it supports is internal to this table. Coverage is 100\% throughout. Bold indicates the setting used in the main analysis, not the minimum within the tested grid. Ranks and ratios quoted in the text are computed from unrounded values, so cells that agree to four decimals need not be equal.}
\label{tab:alpha-k-sweep}
\begin{tabular}{lrrrr}
\toprule
$\alpha$ & $K=25$ & $K=50$ & $K=100$ & $K=200$ \\
\midrule
0.5 & 0.0180 & 0.0182 & 0.0185 & 0.0192 \\
1.0 & 0.0179 & 0.0180 & \textbf{0.0182} & 0.0187 \\
2.0 & 0.0179 & 0.0178 & 0.0178 & 0.0181 \\
\bottomrule
\end{tabular}
\end{table}

Supplementary Table~\ref{tab:alpha-k-sweep} is a single-fold sensitivity analysis rather than a complete cross-validation and characterizes sensitivity within the tested grid without establishing an optimum. Fold-0 MAE spans 8\%, from 0.0178 at $\alpha=2$, $K=50$ to 0.0192 at $\alpha=0.5$, $K=200$. The reported setting ranks ninth among the 12 cells and is 2.7\% above the grid minimum. This analysis was not performed on ELES because the main text's $K=100$ is non-binding for that archive. Although $K=25$ would restrict the largest ELES topology groups, which contain up to 38 states and therefore 37 candidates after query exclusion, an ELES analysis over the same grid would primarily evaluate $\alpha$. Only the smaller $K$ values would constrain the largest groups, so the analysis would not assess the same pool-size--bandwidth trade-off as the BUS39 analysis.

Supplementary Table~\ref{tab:topology-ablation} reports the ELES topology-filter ablation. The column headed $n$ gives the number of matched contingency records, with the number of corresponding states in parentheses. Statistical tests use per-state mean paired differences because contingency records within a state are not independent. States with an exactly zero paired difference are omitted from the Wilcoxon signed-rank calculation, leaving $n_{\neq 0}=222$, 222, and 210 for seeds 42, 123, and 7, respectively. The standardized effect size is $r=z/\sqrt{n_{\neq 0}}$, with the sign of $z$ following the difference defined as filter off minus filter on. A negative value therefore indicates lower error without filtering. Mean paired differences are reported with 95\% bootstrap CIs based on 1,000 resamples.

\begin{sidewaystable}[p]
\centering
\caption{ELES topology-filter ablation across three independent 300-query-state samples. MAE and maximum absolute error are in seconds. Mean paired difference is filter off minus filter on. Off/on excess is the percentage by which the maximum without filtering exceeds the maximum with filtering.}
\label{tab:topology-ablation}
\footnotesize
\begin{tabular}{lrrrrrrrrrr}
\toprule
Seed & $n$ (states) & $W$ & $p$ & $r$ & MAE, on & MAE, off & Mean paired diff. (95\% CI) & Max, on & Max, off & Off/on excess \\
\midrule
42 & 13,856 (224) & 7394.5 & $1.99\times10^{-7}$ & $-0.349$ & 0.0417 & 0.0379 & $-0.0049$ ($-0.0076$, $-0.0024$) & 1.150 & 1.370 & 19.1\% \\
123 & 13,893 (225) & 8068.0 & $6.89\times10^{-6}$ & $-0.302$ & 0.0439 & 0.0406 & $-0.0038$ ($-0.0061$, $-0.0018$) & 1.090 & 1.330 & 22.0\% \\
7 & 12,961 (212) & 8100.0 & $7.32\times10^{-4}$ & $-0.233$ & 0.0394 & 0.0370 & $-0.0025$ ($-0.0044$, $-0.0007$) & 1.320 & 1.380 & 4.5\% \\
\bottomrule
\end{tabular}
\end{sidewaystable}

\begin{table}[htbp]
\centering
\caption{Absolute CCT error quantiles (seconds) for the topology-filter ablation of Supplementary Table~\ref{tab:topology-ablation}, evaluated on the same matched common-support population for each seed. ``On'' denotes exact-topology filtering, and ``off'' denotes no equality filter. The unfiltered condition has lower error at every reported quantile but a higher observed maximum in Supplementary Table~\ref{tab:topology-ablation}.}
\label{tab:topology-ablation-quantiles}
\begin{tabular}{lrrrrrr}
\toprule
& \multicolumn{2}{c}{$q_{90}$} & \multicolumn{2}{c}{$q_{95}$} & \multicolumn{2}{c}{$q_{99}$} \\
\cmidrule(lr){2-3}\cmidrule(lr){4-5}\cmidrule(lr){6-7}
Seed & On & Off & On & Off & On & Off \\
\midrule
42  & 0.120 & 0.110 & 0.190 & 0.171 & 0.460 & 0.410 \\
123 & 0.120 & 0.110 & 0.210 & 0.190 & 0.520 & 0.480 \\
7   & 0.110 & 0.109 & 0.180 & 0.170 & 0.430 & 0.400 \\
\bottomrule
\end{tabular}
\end{table}

\paragraph{Supplementary Note S4: contribution of exact line-status equality.}
None of the 254 Slovenian-specific line-status flags changes state in the ELES record. Consequently, all 1,785 topology
groups result from switching changes in neighboring systems represented by the observability-area model: all 510 of the
907 key flags that vary are outside the assessed system. Neighboring-system switching affects impedance and power flows
within the assessed network, so retaining these flags may be justified. However, the primary analysis did not evaluate
the requirement for exact equality on every flag; this note assesses that requirement.

States are paired when they share a contingency defined by fault location and critical generator, so the two clearing
times correspond to the same conditional fault scenario. Pairs are stratified by the number of line-status flags on
which they differ. Pairs with Hamming distance zero satisfy the deployed key, whereas pairs with nonzero distance are
excluded. Without adjustment for continuous feature distance, the mean absolute clearing-time difference increases
from 0.0586 s within a topology group to 0.1171 s beyond twenty differing flags. This gradient cannot be attributed
solely to switching because mean continuous feature distance increases concurrently. Across strata, the Spearman
correlation of mean clearing-time difference with both flag count and feature distance is 1.000. When generator
commitment is also held constant, pairs differing by one flag have a mean absolute clearing-time difference of 0.0192 s,
compared with 0.0205 s for pairs satisfying the deployed key; the one-flag comparison contains 45 pairs.

Table~\ref{tab:key-locality} compares included and excluded pairs within bands of continuous feature distance, thereby
comparing flag disagreement at approximately similar continuous separation. In the first four bands, which have similar
mean feature distances between conditions, included and excluded pairs have similar mean clearing-time differences. The
fifth band is open-ended and provides a poorer match because excluded pairs have a substantially larger mean feature
distance. Exact line-status equality therefore does not produce demonstrably greater dynamic comparability in this
archive. This result is consistent with the ablation in Tables~\ref{tab:topology-ablation}
and~\ref{tab:topology-ablation-quantiles}, in which disabling the filter reduces error at every reported quantile. The
key remains a compatibility convention that guarantees agreement with the query's switching configuration; it does not
function as a demonstrated accuracy mechanism.

Feature distance in this analysis is Euclidean distance over 5,414 complete standardized continuous columns. It is a
proxy for, but is not identical to, the distance produced by the retrieval scaler. The analysis compares state pairs
rather than complete retrieval pipelines and establishes an association rather than a causal effect.

\begin{table}[htbp]
\centering
\caption{Comparison of included and excluded state pairs within bands of continuous feature distance, ELES. Bands contain
equal numbers of included pairs. ``Included'' pairs agree on all 907 line-status flags and satisfy the deployed key;
``excluded'' pairs differ on at least one flag and do not satisfy the key, regardless of their continuous-feature proximity. Distances are
Euclidean over 5,414 complete standardized continuous columns with topology flags excluded. $|\Delta \mathrm{CCT}|$ is
the mean absolute clearing-time difference over contingencies the two states share. A ratio near one means the key is
not associated with greater dynamic comparability at that separation.}
\label{tab:key-locality}
\begin{tabular}{lrrrrrrr}
\toprule
& \multicolumn{2}{c}{Feature distance} & \multicolumn{2}{c}{$|\Delta \mathrm{CCT}|$ (s)} & & \multicolumn{2}{c}{Contingencies} \\
\cmidrule(lr){2-3}\cmidrule(lr){4-5}\cmidrule(lr){7-8}
Band & Included & Excluded & Included & Excluded & Ratio & Included & Excluded \\
\midrule
1 & 24 & 26 & 0.0278 & 0.0292 & 1.053 & 97,839 & 352 \\
2 & 36 & 38 & 0.0481 & 0.0467 & 0.969 & 83,789 & 1,111 \\
3 & 47 & 48 & 0.0615 & 0.0645 & 1.049 & 76,727 & 5,846 \\
4 & 58 & 60 & 0.0760 & 0.0761 & 1.002 & 70,106 & 33,931 \\
5 & 78 & 102 & 0.0989 & 0.1131 & 1.143 & 64,707 & 1,143,177 \\
\bottomrule
\end{tabular}
\end{table}

\end{document}